\documentclass[aps,prx,reprint,superscriptaddress,amsmath,amssymb,floatfix]{revtex4-2}

\usepackage{graphicx}
\usepackage{physics,tabularray,soul}
\usepackage[dvipsnames]{xcolor}
\usepackage{amsmath,amssymb}
\usepackage{bm}
\usepackage{float}
\usepackage{algorithm}
\usepackage{algpseudocode}
\usepackage[normalem]{ulem}
\usepackage{amsfonts}
\usepackage{mathtools}
\usepackage{tikz}
\usepackage{forest}
\usepackage{makecell}
\usepackage{subcaption}
\usepackage{ragged2e}
\usepackage{hyperref}

\hypersetup{
    colorlinks,
    linkcolor={red!50!black},
    citecolor={blue!50!black},
    urlcolor={blue!80!black}
}
\usetikzlibrary{quantikz2}
\usetikzlibrary{arrows.meta}
\usetikzlibrary{shapes.geometric}
\usetikzlibrary{positioning, fit, backgrounds, calc}
\useforestlibrary{edges}
\UseTblrLibrary{booktabs}

\newcommand{\mzEC}{\gate[style={fill=LimeGreen!50}]{\mathsf{mz_3}}}
\newcommand{\mxEC}{\gate[style={fill=LimeGreen!50}]{\mathsf{mx_3}}}
\newcommand{\mz}{\gate{\mathsf{mz}}}
\newcommand{\mx}{\gate{\mathsf{mx}}}
\newcommand{\rz}{\gate{\mathsf{rz}}}
\newcommand{\rx}{\gate{\mathsf{rx}}}

\newcommand{\leak}[1]{\push{\color{RedOrange}\mathcal{L}}\wire[r][#1][style={RedOrange,dashed}]{q}}
\newcommand{\depol}{\push{\color{blue}\mathcal{D}}}
\newcommand{\errgate}[1]{\gate[style={fill=RedOrange!20,dashed}]{#1}}
\NewDocumentCommand{\cancelgate}{O{2} O{inner sep=-3pt} m}{\gategroup[#1,style={#2,minimum width=0pt,strike out,RedOrange}]{}}

\newcommand{\pleak}{p_\mathrm{leak}}
\newcommand{\pPauli}{p_\mathrm{pauli}}

\DeclareMathOperator*{\argmax}{arg\,max}
\DeclareRobustCommand{\ihat}{\hat{\imath}}
\DeclareRobustCommand{\jhat}{\hat{\jmath}}
\DeclareRobustCommand{\ibhat}{\mathbf{\hat{i}}}
\DeclareRobustCommand{\jbhat}{\mathbf{\hat{j}}}

\makeatletter
\RenewDocumentCommand\eqref{s m}{
  \IfBooleanTF#1%
  {\textup{\tagform@{\ref*{#2}}}}
  {\textup{\tagform@{\ref{#2}}}}
}
\makeatother

\tikzstyle{node} = [rectangle, rounded corners, 
    minimum width=2cm, minimum height=1cm,
    text centered, draw=black, align=center]
\tikzset{slice/.append style={draw=NavyBlue}}

\graphicspath{{Figures/}}

\DeclareCaptionJustification{justified}{\justifying}
\begin{document}

\title{Erasure surface code circuit without mid-circuit erasure checks}

\author{Margaret Pavlovich}
\email{margaret.pavlovich@yale.edu}
\author{Ivan Rojkov}
\affiliation{Yale Quantum Institute, Yale University, New Haven, CT, USA, 06511}
\affiliation{Department of Applied Physics, Yale University, New Haven, CT, USA, 06511}
\author{Chen Wang}
\affiliation{Department of Physics, University of Massachusetts, Amherst, MA, 01003, USA}
\affiliation{Department of Physics, 
University of Toronto, Toronto, ON, M5S 1A7, Canada}
\author{Shruti Puri}
\affiliation{Yale Quantum Institute, Yale University, New Haven, CT, USA, 06511}
\affiliation{Department of Applied Physics, Yale University, New Haven, CT, USA, 06511}
\date{September 4, 2026}

\begin{abstract}

Quantum error correction (QEC) codes can correct twice as many erasure errors as Pauli errors. Because of this scaling advantage, there is significant interest in developing qubits whose dominant error channel can be converted into erasures via mid-circuit erasure checks. 
However, such erasure checks come with hardware overhead in practice. 
End-of-the-line three-state readout,  in which one simultaneously measures a qubit's erasure status and computational state, is an alternative to mid-circuit erasure checks that is generally simpler to implement.
In this work, we systematically study the conditions required to enable erasure performance---the doubled error-correction capacity---in the surface code with and without mid-circuit erasure checks.
We introduce the \emph{moonwalking surface code}, the time-reversal of the walking surface code, as a zero-overhead circuit with superior handling of leakage and erasure. 
Specifically, we show that it enables erasure-like logical error rate scaling when combined with three-state measurement if leaked qubits cause two-qubit gates to be skipped and an appropriate decoder is used. 
Our decoder, based on a branch-and-bound algorithm, specifically incorporates the noise structure of the skip-gate leaked-qubit effect.  

\end{abstract}

\maketitle

\section{Introduction}

Quantum computers will require quantum error correction (QEC) to perform useful large-scale computation.
In the context of QEC codes, there exists a hierarchy of errors from hardest to easiest to correct. 
At the bottom of the hierarchy (hardest to correct) are leakage errors, where a physical qubit may undetectably leave its logical subspace. 
Stochastic Pauli errors, in which a qubit may randomly undergo a bit and/or phase flip, are in the middle of the hierarchy. 
At the top of the hierarchy (easiest to correct) are erasure errors, in which a qubit is known to have effectively undergone a Pauli error at a particular circuit location, but precisely which error occurred is unknown \cite{bennettCapacitiesQuantumErasure1997,grasslCodesQuantumErasure1997a}. 
In general, QEC codes can correct twice as many erasure errors as Pauli errors~\cite{bennettCapacitiesQuantumErasure1997,grasslCodesQuantumErasure1997a}.
Because of this scaling advantage, there is interest in developing \emph{erasure qubits}---combinations of hardware and qubit encoding whose dominant noise can be converted into erasure errors---and QEC circuits which can take advantage of them without significant overhead. 

\begin{table*}[!ht]
    \centering
    \begin{tblr}{colspec={l@{}cX[3]X[1]},rowspec={Q[b]*{4}{Q[m]}}}
    \toprule
    Leaked-qubit effect & Diagram & Description & Motivation \\ \midrule
    Depolarizing & 
    \begin{quantikz}[column sep=6pt, row sep=6pt]
        \lstick{\color{RedOrange}$\mathcal{L}$}
        &  \ctrl{1} \setwiretype{n}
        & \wire[l][2][style={RedOrange,dashed}]{q} \\
        & \targ{}
        & \rstick{\color{RedOrange}$\mathcal{D}$}
    \end{quantikz} 
    &
    The other qubit undergoes the depolarizing error channel $\mathcal{D}=\mathcal{E}\{I,X,Y,Z\}$. 
    &
    Worst-case baseline \cite{fowlerCopingQubitLeakage2013,guFaulttolerantQuantumArchitectures2025}
    \\
    Tailored & 
    \begin{quantikz}[column sep=6pt, row sep=6pt]
        \lstick{\color{RedOrange}$\mathcal{L}$} 
        &  \ctrl{1}\setwiretype{n}
        & \wire[l][2][style={RedOrange,dashed}]{q} \\
        & \targ{}
        & \rstick{\color{RedOrange}$\mathcal{E}\{I,X\}$}
    \end{quantikz}
    \&
    \begin{quantikz}[column sep=6pt, row sep=6pt]
        &  \ctrl{1}
        & \rstick{\color{RedOrange}$\mathcal{E}\{I,Z\}$}
        \\
        \lstick{\color{RedOrange}$\mathcal{L}$} 
        & \targ{}\setwiretype{n}
        & \wire[l][2][style={RedOrange,dashed}]{q}
    \end{quantikz}
    &
    The other qubit undergoes the $\mathcal{E}\{I,X\}$ or $\mathcal{E}\{I,Z\}$ error channel when it is the target or control of the gate, respectively.
    & 
    Trapped ion qubits \cite{brownLeakageMitigationQuantum2019}
    \\
    Skip-gate & 
    \begin{quantikz}[column sep=8pt, row sep=6pt]
        \lstick{\color{RedOrange}$\mathcal{L}$} 
        & \ctrl{1}\cancelgate{}\setwiretype{n} 
        & \wire[l][2][style={RedOrange,dashed}]{q}\\
        & \targ{}
        & 
    \end{quantikz}
    &
    The two-qubit gate is skipped/cancelled, indicated by the struck-out CX gate symbol. This is deterministic.
    &
    Dual-rail qubits \cite{mehtaBiaspreservingErrordetectableEntangling2025a}
    \\ \bottomrule
    \end{tblr}
    \caption{
    Models considered for the effect of leaked qubits interacting with non-leaked qubits, presented in order of least to most structure. 
    $\mathcal{L}$ represents the leakage channel and the qubit wire is dashed and colored red-orange while it is leaked. 
    $\mathcal{E}\{\cdot\}$ is the channel which applies each element of $\{\cdot\}$ with equal probability.
    }
    \label{tab:leakage-effects}
\end{table*}

Typically, erasure qubits are constructed by encoding the qubit such that dominant hardware noise causes leakage; this leakage is then detected without disturbing the qubit's computational subspace, thus converting the dominant noise into erasures. 
Such leakage checks, which remarkably connect the two ends of the noise hierarchy, are known as \textit{mid-circuit erasure checks}. 
When applied during the circuit execution, they provide fine-grained spacetime information about leakage events which guarantees the erasure-scaling advantage.
Erasure qubits with mid-circuit erasure checks have been proposed and demonstrated in neutral atoms~\cite{wuErasureConversionFaulttolerant2022,schollErasureConversionHighfidelity2023,maHighfidelityGatesMidcircuit2023}, trapped ions~\cite{kangQuantumErrorCorrection2023,quinnHighfidelityEntanglementMetastable2026},
and superconducting hardware~\cite{kubicaErasureQubitsOvercoming2023a,teohDualrailEncodingSuperconducting2023,levineDemonstratingLongCoherenceDualRail2024,koottandavidaErasureDetectionDualRail2024,liuHardwareEfficientErasureQubits2026}.
Nonetheless, mid-circuit checks are experimentally challenging to implement and they may themselves be noisy and impose a circuit depth penalty.
We propose to avoid these costs by relying solely on three-state readout (also referred to as state-selective readout) rather than mid-circuit erasure checks. 
A three-state readout returns one of three outcomes corresponding to either the erased, the computational 0 or the computational 1 states~\cite{KLM2001,wu2019stern,bluvstein2026fault,neeley2009emulation,bianchetti2010control,chen2023transmon}.
Intuitively, a three-state readout can be thought of as a mid-circuit erasure check followed immediately by a conventional computational-basis measurement. 

In this work, we investigate the requirements on qubit noise to maintain erasure scaling when mid-circuit erasure checks are reduced or eliminated. 
Remarkably, when a code is appropriately implemented, we find that the erasure scaling can be maintained with only three-state readouts if leaked qubits cause two-qubit gates in which they participate to be skipped. 
We introduce an implementation of the surface code referred to as the \emph{moonwalking surface code}, which achieves erasure performance with zero overhead when augmented with three-state readout, skip-gate leakage, and a decoder to leverage the specific noise structure of skip-gate leakage.

We note that prior work has studied the impact of erasure check frequency in Ref.~\cite{guFaulttolerantQuantumArchitectures2025}, but considered only one model for the errors introduced when leaked qubits interact with other qubit during two-qubit gates. 
The impact of different leaked-qubit effects has been studied in Refs.~\cite{changSurfaceCodeImperfect2025,brownLeakageMitigationQuantum2019}, but not alongside varied erasure check schedules, and their analysis did not include the skip-gate leaked-qubit effect. 
Similar circuits to our moonwalking surface code with three-state readout have recently been considered for application to neutral atom quantum computing \cite{yuTamingRydbergDecay2026a,liuAchievingOptimalDistanceAtomLoss2026}.
Compared to previous works, we thoroughly study how erasure check frequency and leaked-qubit effect may compensate for one another. 

Our paper is structured as follows. 
In Sec.~\ref{sec:qubit-design-space}, we describe the two properties of erasure qubits whose impact we analyze: the effect of leaked qubits in two-qubit gates (Sec.~\ref{sec:leaked-qubit-effects}) and the type of erasure check available (Sec.~\ref{sec:ec-types}). 
In Sec.~\ref{sec:circuit-space}, we show the syndrome extraction circuits used in this work, including the moonwalking surface code (Sec.~\ref{sec:circuits}), and explain how erasure checks are integrated into these circuits (Sec.~\ref{sec:ec-sched}). 
We introduce our branch-and-bound decoder in Sec.~\ref{sec:decoding}. 
Finally, our QEC simulation results are presented in Sec.~\ref{sec:qec-results}, including predicted asymptotic scaling (Sec.~\ref{sec:min-weights}) and observed logical error rates (Sec.~\ref{sec:LER}).

\section{Erasure qubit design space}
\label{sec:qubit-design-space}

In this section, we describe our models of leakage and its effects, and two types of erasure checks. 
Together, the effects of leakage and the type(s) of erasure check available define a design space for erasure qubits. We will show in Sec.~\ref{sec:qec-results} that these two aspects of erasure qubit design may compensate for one another.

We assume that leakage is equally likely from the $\ket{1}$ state as the $\ket{0}$ state. 
This can be arranged in design (such as matching decay from the two modes of a dual-rail qubit) or by twirling~\cite{wallmanNoiseTailoringScalable2016}. 

\subsection{Leaked-qubit effects}
\label{sec:leaked-qubit-effects}

Next, we consider what happens when a leaked qubit interacts with a non-leaked qubit during a two qubit gate. The three two-qubit gate noise models we consider are {\it depolarizing leakage, tailored leakage}, and {\it skip-gate leakage}, described below and summarized in Table~\ref{tab:leakage-effects}. Our circuits use CX (also known as CNOT) as their two-qubit gate. 

In both depolarizing and tailored leakage models, a leaked qubit induces a stochastic Pauli error on the qubit that it interacts with. 
In the depolarizing model, the Pauli operator is drawn uniformly from $\{I,X,Y,Z\}$~\cite{fowlerCopingQubitLeakage2013,ghosh2013understanding,suchara2015leakage,brown2018comparing,brown2019leakage,guFaulttolerantQuantumArchitectures2025,changSurfaceCodeImperfect2025}. 
In the tailored model, the Pauli error is drawn uniformly from $\{I,Z\}$ if non-leaked qubit is the control of the CX gate, or from $\{I,X\}$ if the non-leaked qubit is the target~\cite{changSurfaceCodeImperfect2025}. 

Finally, in the skip-gate leakage model, any gate a leaked qubit is to participate in does not happen, i.e., the gate is skipped. 
In this model, when a qubit leaks during a gate operation, it is equivalent to the leakage occurring before or after the gate; whether any given mid-gate leakage event behaves as a before-gate or after-gate leakage may be stochastic. 
Notably, the noise model described in Refs.~\cite{wuErasureConversionFaulttolerant2022,yuTamingRydbergDecay2026a} for leakage errors in neutral atom erasure qubits does not have this property. 
As described in those works, leakage during the gate can cause a phase error on the qubit it was interacting with, though subsequent gates are skipped. 
We found that this leaked-qubit effect model cannot reach the optimal erasure scaling with only three-state measurement using any of the circuits we considered, and thus do not include the results in this paper.
Further analysis is required to determine whether the true skip-gate leaked-qubit effect is achievable in neutral atom qubits. 

\subsection{Types of erasure check}\label{sec:ec-types}

In the erasure check design space, we consider ideal mid-circuit erasure checks and ideal three-state readouts. A mid-circuit erasure check discriminates between the leakage state, $\ket{\mathrm{e}}$, and the computational subspace with $100\%$ fidelity while preserving computational-basis information. If the qubit is found to be in $\ket{\mathrm{e}}$ then it is returned to the computational subspace in a completely mixed state. In practice, mid-circuit erasure checks can have false positives and negatives and the effects of such imperfections have been studied previously~\cite{changSurfaceCodeImperfect2025,guOptimizingQuantumErrorCorrection2025}. A three-state measurement discriminates between all three states with $100\%$ fidelity. In this work, we will always reset the measured qubit in a computational basis state after the 3-state measurement. In practice, the three-state readout can also be faulty and the impact of such faults will be explored in future work.

\begin{table*}[!t]
    \centering
    \begin{tblr}{
        colspec={cccc},colsep=4pt,rowspec={Q[f]Q[f]},
        column{1}={colsep=0pt},column{4}={rightsep=0pt}}
    \toprule
    $Z$ Plaquette &
    (Static) surface code & Walking surface code & Moonwalking surface code \\\midrule
    \begin{tikzpicture}[baseline={([yshift=-2pt]C)}]
        \coordinate (tl) at (0,18pt);
        \coordinate (tr) at (1,36pt);
        \coordinate (br) at (1,-18pt);
        \coordinate (bl) at (0,-36pt);
        
        \draw[blue, thin, fill=blue!20] (tl) -- (tr) -- (br) -- (bl) -- cycle;

        \fill[black] (tl) circle (2pt);
        \fill[black] (tr) circle (2pt);
        \fill[black] (br) circle (2pt);
        \fill[black] (bl) circle (2pt);
        
        \coordinate (C) at ($ (tl)!0.5!(br) $); 
        \fill[black] (C) circle (2pt);
        
        \node[below right] at (tr) {1};
        \node[below left]  at (tl) {2};
        \node[above right] at (br) {3};
        \node[above left]  at (bl) {4};
        
        \node[above right] at (C) {$a$};

        \coordinate (Xend) at (1.5,0);        
        \draw[black, thick] (tl) -- (tl -| Xend);
        \draw[black, thick] (tr) -- (tr -| Xend);
        \draw[black, thick] (br) -- (br -| Xend);
        \draw[black, thick] (bl) -- (bl -| Xend);
        \draw[black, thick] (C) -- (C -| Xend);
    \end{tikzpicture}
    &
    \begin{quantikz}[row sep={18pt,between origins},column sep=12pt,align equals at=3]
            & \targX[style=violet]{}\wire[d][2]["\text{\footnotesize{L}}"{right, pos=0.25},violet,dotted]{a}
                                    & \ctrl{2}\slice{A}&\slice{B}&\slice{C}&\slice{D}& \\
            &                       &                  & \ctrl{1}&         &         & \\
        \rz & \targX[style=violet]{}& \targ{}          & \targ{} & \targ{} & \targ{} & \mz \\
            &                       &                  &         &\ctrl{-1}&         & \\
            &                       &                  &         &         &\ctrl{-2}& 
    \end{quantikz}
    &
    \begin{quantikz}[row sep={18pt,between origins},column sep=12pt,align equals at=3]
            & \ctrl{2}\slice{A}&\slice{B}&\slice{C}&\slice{D}&  \mz \\
            &                  & \ctrl{1}&         &         &  \mx  \\
        \rz & \targ{}          & \targ{} & \targ{} &\ctrl{2}&        \\
            &                  &         &\ctrl{-1}&         &  \mx  \\
            &                  &         &         & \targ{} &  \mz 
    \end{quantikz}
    &
    \begin{quantikz}[row sep={18pt,between origins},column sep=12pt,align equals at=3]
           & \ctrl{2} &          &          &                 & \mz \\
           &          &          & \ctrl{1} &                 & \mx \\
        \rz& \targ{}  & \targ{}  & \targ{}  & \ctrl{2}        &     \\
           &          & \ctrl{-1}&          &                 & \mx \\
           & \slice{A}& \slice{B}& \slice{C}& \targ{}\slice{D}& \mz 
    \end{quantikz}
    \\ \bottomrule
    \end{tblr}
    \caption{{
    One round of $Z$-stabilizer measurement for each of the three circuits analyzed (Sec.~\ref{sec:circuits}), with markers we will use to define the four erasure check (EC) schedules (Sec.~\ref{sec:ec-sched}). 
    The $\mathsf{rz}$ gate resets the qubit into the $\ket{0}$ state.
    The $\mathsf{mz}$ and $\mathsf{mx}$ gates measure the qubit in the $Z$ and $X$ basis, respectively. 
    For the static surface code, the ancilla (qubit $a$) measurement result gives the $Z$ stabilizer value, while for the walking and moonwalking circuits, the stabilizer is measured by qubit 4. 
    The violet, dotted SWAP gate labeled ``L'' indicates a leakage-SWAP gate which leaves the computational-basis states untouched; it is only used for EC schedule 8 in the static surface code circuit. 
    For EC schedule 8 in all circuits, every measurement is replaced by three-state measurement. 
    The remaining EC schedules use mid-circuit erasure checks: 
    EC schedule 4 has erasure checks on all qubits at time slice D; 
    EC schedule 2 has erasure checks at time slices B and D; and 
    EC schedule 1 has erasure checks at all time slices A, B, C, and D. 
    When a mid-circuit erasure check is followed immediately by a measurement, they may be combined into a three-state measurement. 
    }}
    \label{tab:circuits}
\end{table*}

When three-state readout is performed regularly on only a subset of qubits, it is necessary to use leakage-SWAP gates to enable erasure conversion on all qubits without mid-circuit erasure checks. The leakage-SWAP gate swaps the computational $\ket{0}$ state with the leakage state $\ket{\mathrm{e}}$. It is always applied between an ancilla and data qubit immediately after the ancilla is reset to $\ket{0}$ \footnote{If the ancilla is to be reset to $\ket{+}$, we suppose it is reset to $\ket{0}$, then the leakage-SWAP gate is applied, then a Hadamard is applied to the ancilla qubit.}, transferring leakage from data to ancilla qubits and resetting any leaked data qubits to the qubit subspace. Where leakage-SWAP gates are used (see Sec.~\ref{sec:circuits}), they are considered noiseless.

\section{Circuit Design Space}\label{sec:circuit-space}

\begin{figure*}
    \centering
    \includegraphics{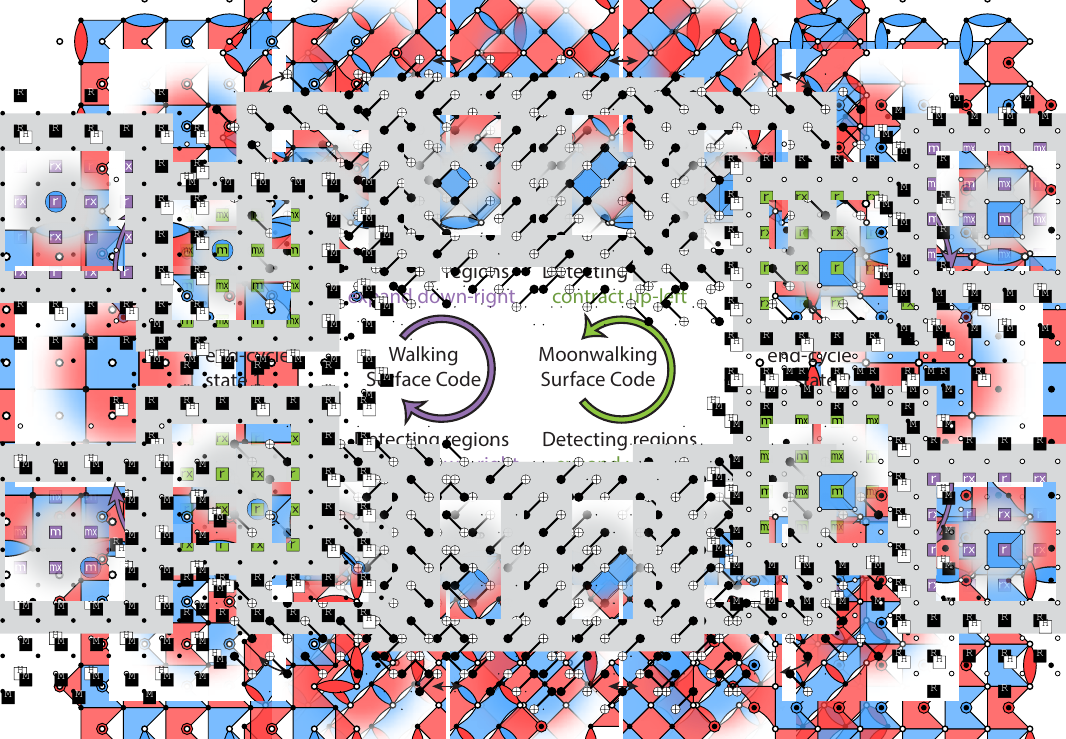}
    \caption{
    The moonwalking surface code is the time-reverse of the original walking surface code. Here, we show the detecting regions and CX gates (illustrations adapted from those generated by Stim \cite{stim}) for two syndrome extraction cycles of each circuit.
    Qubits are depicted as alternating white and black dots to make plaquette shifts more apparent. 
    $X$-type detecting regions are red and $Z$-type detecting regions are blue. 
    The walking surface code circuit follows the loop clockwise and uses the outer, purple operations for measurements and resets.     
    The moonwalking surface code circuit follows the loop counterclockwise and uses the inner, green operations for measurements and resets. 
    Both circuits use the same CXs. 
    The CX gates transform the overlaid detecting region between the two states shown. Which shape represents the initial state of the detecting region and which the final depends on the direction we are traversing the loop, i.e., whether we are considering the walking or moonwalking surface code. 
    }
    \label{fig:moon-orig-reverse}
\end{figure*}

In this section we introduce the moonwalking surface code, 
which we will compare to the conventional static and walking surface code implementations in terms of their performance against leakage and erasure errors in Sec.~\ref{sec:qec-results}. 
Leakage-to-erasure conversion is integrated into all three circuits using mid-circuit erasure checks and/or three-state measurements, plus leakage-SWAP gates when necessary. 

\subsection{Syndrome extraction circuits}\label{sec:circuits}

The static, walking, and moonwalking surface code circuits differ in their syndrome extraction circuits as seen from the $Z$ stabilizer measurement circuit snippets shown in Table~\ref{tab:circuits}.

The static surface code uses the standard circuit in which CX gates are applied between an ancilla initialized in $\ket{0}$ ($\ket{+}$) and the data qubits of a $Z$ ($X$) plaquette and then the ancilla is measured in the $Z$ ($X$) basis. 
As will be discussed in Sec.~\ref{sec:ec-sched}, we cannot convert every qubit's leakage into erasure in the static surface code circuit without mid-circuit erasure checks or leakage-SWAP gates. 
For this reason, we are interested in dynamic circuits which reassign data and ancilla qubits every syndrome extraction cycle. This reassignment means every qubit is regularly measured and reset, enabling erasure conversion with only three-state measurement. 

The specific dynamic circuits we study are the walking surface code, introduced in Ref.~\cite{mcewenRelaxingHardwareRequirements2023}, and the proposed moonwalking surface code. 
Remarkably, we find that even though the moonwalking circuit is simply the time-reversed walking circuit, it has significantly improved scaling when using infrequent erasure checks under skip-gate leakage. We note that the existence of a time-reversed walking circuit was mentioned in Ref.~\cite{mcewenRelaxingHardwareRequirements2023} but its unique properties, particularly regarding different leakage noise effects and including erasure conversion, were not studied.

\begin{figure}[htb]
    \centering
    \begin{quantikz}[column sep=4pt, row sep={16pt,between origins}, align equals at=1.5]
        &    &[2pt] \ctrl{1}& \\
        & \rz&      \targ{} & 
    \end{quantikz}
    $\!=\!$
    \begin{quantikz}[column sep=6pt, row sep={16pt,between origins}, align equals at=1.5]
        &    &[2pt] \swap{1} &[6pt] \ctrl{1}\gategroup[2,steps=3,style={dashed,inner sep=1pt}]{SWAP} & \targ{} & \ctrl{1} &[2pt] \ctrl{1} & \\
        & \rz& \targX{} & \targ{}  & \ctrl{-1} & \targ{}  & \targ{}  &
    \end{quantikz}
    $\!=\!$
    \begin{quantikz}[column sep=6pt, row sep={16pt,between origins}, align equals at=1.5]
        &[2pt] \swap{1} &[2pt] \rz &[-2pt] \ctrl{1} & \targ{}   & \\
        &      \targX{} &          &       \targ{}  & \ctrl{-1} & 
    \end{quantikz}
    $\!=\!$
    \begin{quantikz}[column sep=6pt, row sep={16pt,between origins}, align equals at=1.5]
        & \swap{1} &[2pt] \rz & \targ{} &[-2pt] \\
        & \targX{} &          &\ctrl{-1}& 
    \end{quantikz}
    \caption{
    The circuit equivalences and transformations to construct the moonwalking surface code circuit. 
    A pair of SWAP operations is inserted between the reset operation and first CX gate. 
    One SWAP is commuted through the reset operation to the beginning of the syndrome extraction circuit. 
    In the context of the whole code, this SWAP is meaningless. Its effect can be recovered by simultaneously shifting all the data qubits along a diagonal by half a plaquette in software so that they appear to be placed on the sublattice formed by ancilla qubits of the previous round.
    The other SWAP is decomposed into CX operations which are combined with the existing CX gate and the reset operation, which has the cumulative effect of flipping the original CX gate.
    }
    \label{fig:moonwalking-construction}
\end{figure}
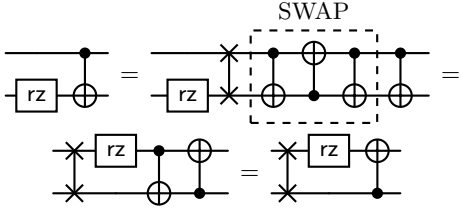

The walking and moonwalking surface code circuits can each be constructed from the static surface code circuit by inserting a pair of SWAP operations. 
For the moonwalking circuit, the SWAPs are inserted between the reset and the first CX, as shown in Fig.~\ref{fig:moonwalking-construction}. 
For the walking circuit, the SWAPs are inserted between the last CX and the measurement, as shown in Fig.~\ref{fig:walking-construction}. 
One of the SWAPs is commuted through the reset (measurement) for the moonwalking (walking) circuit. 
In the context of the code, the effect of the SWAP between syndrome extraction rounds is trivial as it can be recovered by shifting the data qubits diagonally by half a plaquette in software so that they appear to be placed on the ancilla qubit sublattice of the previous round. Indeed, the simple removal of the SWAP still gives a good syndrome extraction circuit. 
These circuits move the logical patch one step diagonally. By spatially rotating the circuit for the next round of syndrome extraction, the patch is stepped diagonally back to its original position. 
To finish constructing the moonwalking (walking) circuit, the remaining SWAP is decomposed into CX gates which are combined with the existing CX gate and the reset (measurement) operation, which allows us to recover a syndrome extraction circuit with the same depth as the original. 
The $Z$ syndrome extraction circuits are shown in Table~\ref{tab:circuits}.

We observe from Fig.~\ref{fig:moon-orig-reverse} that the gate sequence of moonwalking circuit is the time-reverse of the walking surface code circuit---hence our choice of the name ``moonwalking.''
Finally, we note that extra qubits ($\sim\ell$ extra qubits for an $\ell\times\ell$ code) must be added at the boundaries of the surface code patch to give the logical patch somewhere to move to and to maintain gate locality, and care must be taken to ensure that added gates do not reduce the Pauli distance of the code. 
This is the same for both the moonwalking and walking circuits. 
See Supplementary Materials for details on gates and qubits at the boundaries.

We note that the moonwalking circuit could be equivalently obtained from the mid-SWAP circuit in Ref.~\cite{liuAchievingOptimalDistanceAtomLoss2026} (see Fig.~\ref{fig:mid-SWAP}). 
However, there are some important differences between mid-SWAP circuit as implemented in ~\cite{liuAchievingOptimalDistanceAtomLoss2026} and the moonwalking circuit. The former is based on a real SWAP between the data and ancilla qubits implemented by qubit shuttling. Such shuttling operations, avoided in the moonwalking surface code, are not possible in every qubit platform without additional hardware costs. 
Moreover, in Ref.~\cite{liuAchievingOptimalDistanceAtomLoss2026} boundary qubits are only measured out every four syndrome extraction cycles. In contrast, the moonwalking circuit does not require any real SWAP operations and every qubit is measured out every other syndrome extraction cycle.

\subsection{Circuits enhanced with leakage-to-erasure conversion}\label{sec:ec-sched}

Here, we describe our four different protocols for converting leakage into erasure, which we call \emph{erasure check (EC) schedules}. 
Each of the three syndrome extraction circuits (static, walking, and moonwalking) can be combined with any of the EC schedules, as shown in Table~\ref{tab:circuits}.

We label the EC schedules by the number of two-qubit gates executed between sequential erasure checks, i.e., the maximum number of two-qubit gates a leaked qubit may affect before the leakage is converted into erasure and removed.
Three of the EC schedules use mid-circuit erasure checks while the fourth uses only three-state measurement (see Sec.~\ref{sec:ec-types}). 
The first EC schedule is schedule 1, for which we insert mid-circuit erasure checks after every two-qubit gate of a syndrome-extraction circuit.
Next, we consider inserting mid-circuit erasure checks after the second and fourth two-qubit gates of a syndrome-extraction circuit, which gives EC schedule 2, or after only the fourth two-qubit gate, which gives EC schedule 4. 
These three EC schedules were previously studied in Ref.~\cite{guOptimizingQuantumErrorCorrection2025}, though only under the depolarizing leaked-qubit effect~\footnote{Note that our EC schedule labels are reversed relative to Ref.~\cite{guOptimizingQuantumErrorCorrection2025}, as ours label period, whereas theirs label the frequency}.
Finally, we consider the possibility of replacing each measurement in the syndrome extraction circuit with three-state readout while using no mid-circuit erasure checks, which we call EC schedule 8. 
In the case of the static circuit, a leakage-SWAP gate is applied between each data qubit and a neighboring ancilla before the first two-qubit gate in EC schedule 8 (see Table~\ref{tab:circuits}). 

\section{Decoding}\label{sec:decoding}

In this section we present the challenges and our solutions to the decoding problem in the presence of leakage and erasure. When leakage is detected in any of the erasure check schedules described in the previous section, we know which qubits leaked since the previous erasure check, however, the time-step or circuit location of the is leakage unknown. 
Importantly, since a qubit may only leak once, the leakage location is described with a disjoint probability distribution, not by independent probabilities of leakage at each time step. 
A disjoint probability distribution is one where the events are mutually exclusive: one event in the distribution occurring precludes any other event from occurring. In other words, the events described by a disjoint probability distribution are maximally anti-correlated. 
As we will describe below, this disjointness changes the nature of the decoding problem. 
First, we outline how standard surface code decoders can be calibrated to handle the effect of leakage errors.

Surface code decoders rely on the syndrome extraction circuit's detector error model (DEM) which represents a collection of elementary errors that may occur in a circuit, their associated probabilities of occurrence, and the stabilizers and logical outcomes they flip. 
This DEM is used to generate a graph for minimum-weight perfect matching (MWPM). 

From the perspective of calibrating the decoder, it is beneficial to have an analytical description of the DEM (see Appendix~\ref{app:dems-graphs} for more details). Given that such analytical descriptions are easily obtained for stochastic Pauli errors, it is convenient to describe the effects of leakage in a circuit through a Pauli error model. 
A Pauli error model whose effects in the circuit and measurement statistics \emph{contains} the
effects of a leakage event is called its \emph{Pauli envelope} \cite{liuAchievingOptimalDistanceAtomLoss2026}. 
Specifically, a leakage event does not deterministically flip certain detectors and logical outcomes; instead, it is associated with a (uniform) distribution of such outcomes. 
A stochastic Pauli channel which can give rise to those same outcomes is a Pauli envelope for the leakage event. 
It may not be possible to define a stochastic Pauli channel whose possible outcomes include only the possible outcomes of the leakage event. 
Instead, the set of possible outcomes may be overcomplete; this is why this channel is called an ``envelope.''

The Pauli envelope for a given leakage event depends on the circuit location of the leakage event, the leaked-qubit effect model, and the EC schedule. The Pauli envelope under the skip-gate leaked-qubit effect has been described previously~\cite{liuAchievingOptimalDistanceAtomLoss2026}. We describe all the Pauli envelopes relevant for this paper in Appendix~\ref{app:pauli-envelope}. 
Importantly, for the depolarizing and tailored leaked-qubit effects, the Pauli envelope for a qubit leakage at a specific time step is equivalent to the composition of the
Pauli envelopes of leakage applied to that qubit at all subsequent time steps until it is reset. 
In other words, the Pauli envelopes are cumulative. 
Because of this, the individual Pauli errors in the Pauli envelopes may be treated as independently occurring events without affecting the circuit's fault-tolerance against leakage, even though the leakage events are themselves disjoint. 
Therefore, for these leaked-qubit effects, we can accurately decode using a DEM generated from the marginal probabilities of errors calculated from the Pauli envelopes of all leakage events that could have triggered each positive erasure check. 
As in Ref.~\cite{liuAchievingOptimalDistanceAtomLoss2026}, we call this approach \emph{marginal decoding}, and note that it is equivalent to the method used in Ref.~\cite{guFaulttolerantQuantumArchitectures2025}.

On the other hand, for all but EC 1, the Pauli envelopes for leakage events under the skip-gate leaked-qubit effect are not cumulative. 
A Pauli envelope under this leaked-qubit effect comprises a full depolarization on the leaked qubit at the time of the leakage plus at a few, but not all, subsequent time steps (see Appendix~\ref{app:pauli-envelope}).
Therefore, the individual Pauli errors which make up the Pauli envelopes must be treated as disjoint similarly to the leakage events themselves, and thus these Pauli errors cannot be represented by a single DEM (see Appendix~\ref{app:dems-graphs}). 
This means we cannot rely solely on standard surface code decoding techniques. 

In principle, for the skip-gate leaked-qubit effect, the decoder could generate a DEM using the Pauli envelopes for each possible combination of leakage events that could have triggered the observed positive erasure checks. 
Then, one could perform MWPM on the graph associated with each DEM, and choose the lowest-weight matching for an accurate decoding result; see Appendix~\ref{app:brute-force-mle} for details. 
The number of such DEMs grow exponentially with the code size and this brute-force method, although accurate, is clearly not scalable. 
At the other extreme, an efficient but inaccurate approach is to use the marginal decoding strategy that uses the marginal probability of Pauli errors based on the Pauli envelopes. 
The marginal strategy approximately enforces the disjointness of leakage locations, because Pauli errors which are consistent with multiple leakage locations are assigned a higher probability than those which are consistent with fewer. 

In this work we take a hierarchical approach for accurate decoding that is faster on average than the brute-force approach. 
We call this approach \emph{branch-and-bound (B\&B) decoding}, which begins by performing the marginal decoding. 
Then, we check if this solution is consistent with the disjointness of leakage locations. In the vast majority of cases ($>99\%$ in the parameter regime examined), the marginal solution is valid in this sense. 
If so, we accept the marginal solution. 
If the marginal solution is not valid, we pick one erasure check and strictly enforce its leakage location disjointness. 
We iteratively enforce more and more erasure checks' disjointness until a valid solution is found. See Appendix~\ref{app:branch-bound-mle} for details.  
We note that this strategy may also be used when leakage errors are accompanied by stochastic Pauli errors.
The AI assistant Claude Code was used in algorithm design and implementation for the branch-and-bound search, implemented as a priority queue, and the validity-checking subroutine. 

\section{Results}\label{sec:qec-results}

In this section, we present results from circuit-level memory simulations for each syndrome extraction circuit (Sec.~\ref{sec:circuits}), erasure check schedule (Sec.~\ref{sec:ec-sched}), and leaked-qubit effect (Sec.~\ref{sec:leaked-qubit-effects}).
We implement the decoders discussed in Sec.~\ref{sec:decoding}, using PyMatching~\cite{pymatching} for the minimum-weight perfect matching subroutine. 
The QEC circuits are analyzed and sampled using Stim~\cite{stim}. 
Our code for generating and sampling circuits with leakage and for decoding with infrequent erasure checks is available on GitHub \cite{githubSCLE}.

In Sec.~\ref{sec:min-weights}, we find the minimum uncorrectable error weight for each design choice. 
In Sec.~\ref{sec:LER}, we find the threshold and the logical error rate (LER) scaling for each design choice, and find agreement with our predictions based on the minimum uncorrectable error weights. 
Importantly, we find that the minimum uncorrectable error weight is $\ell$ for an $\ell\times\ell$ moonwalking surface code with three state measurement under the skip-gate leaked-qubit effect when using branch-and-bound decoding.

\subsection{Minimum fault weights}\label{sec:min-weights}

The asymptotic scaling of the logical error rate for a QEC circuit can be predicted by determining the minimum number of physical errors required to cause a logical error, known as the fault distance. 

First, we study the leakage fault distance $d_\mathrm{L}$, which we define as the minimum number of leakage events which can cause an \emph{undetectable} logical error in a given QEC circuit. 
For errors to be undetectable, erasure check information is discarded. However, mid-circuit erasure checks and three-state measurement still reset leaked qubits to the computational subspace.
Therefore, the EC schedule determines how often leakage is removed from the circuit, and may affect the leakage distance even though erasure check information is not used.
Specifically, more frequent leakage removal restricts the size of each leakage event's Pauli envelope and can improve the leakage distance.

We determine the leakage fault distance's dependence on code size for each combination of syndrome extraction circuit, EC schedule, and leaked-qubit effect based on the Pauli envelopes of possible leakage events. 
Starting with an elementary space-time tile of the standard Pauli detector graph (i.e., the DEM), we determine which edges correspond to a given leakage event's Pauli envelope. 
Most edges are in multiple leakage events' Pauli envelopes. 
We then determine the minimum number of unique leakage events' Pauli envelopes required to traverse the detector graph in the boundary-to-boundary direction. 
The ratio of this minimum number of leakage events to the ordinary Pauli distance, $\ell$, determines how the leakage distance, $d_\mathrm{L}$, scales with the surface code size. 
The boundaries of the detector graph are not considered in this analysis, and therefore we determine only the scaling of $d_\mathrm{L}$ with $\ell$, and neglect any constant terms. 
The leakage distances are presented in Table~\ref{tab:leakage-d}.

\begin{table}[htb]
    \centering
    \begin{tblr}{@{}lccc@{}}
    \toprule
    \SetCell[r=2]{l,c}{{Leaked-qubit\\effect model}} & \SetCell[c=3]{c}{Erasure check schedule} \\ \cmidrule{2-4}
                 &  1      &  2                     &  4 or 8     \\ \midrule
    Depolarizing &  $\ell$ &  $\frac{1}{2}\ell$     &  $\frac{1}{2}\ell$   \\
    Tailored     &  $\ell$ &  $\ell$                &  $\frac{1}{2}\ell$   \\
    Skip-gate    &  $\ell$ &  $\ell$                &  $\ell$    \\
    \bottomrule
    \end{tblr}
    \caption{
        Scaling of leakage fault distance, $d_\mathrm{L}$, for an $\ell\times\ell$ surface code, which has Pauli distance $\ell$. The leakage fault distance is the minimum number of leakage events which can cause an undetectable logical error in the absence of erasure information. 
        The erasure check schedule here functions as a leakage removal schedule. 
        Notice the overall trend that less-structured leaked-qubit effects require more frequent leakage removal to avoid harming code distance, i.e., to have $d_\mathrm{L}=\ell$. 
        All three circuits studied here---the static, walking, and moonwalking surface code---have the same leakage fault distance. 
    }
    \label{tab:leakage-d}
\end{table}

The leakage fault distance sets an upper bound on the number of leakage errors that could be corrected when erasure information is included. 
Specifically, if $d_\mathrm{L}$ leakage events can cause an undetectable logical error, then up to $d_\mathrm{L}-1$ leakage events can be reliably corrected with perfect information about which leakage events happened, even without knowing which Pauli errors they effectively led to. 
However, infrequent erasure checks, as studied in this work, provide only partial information about the leakage events. 
We must therefore separately assess the correctability of leakage errors given such imprecise erasure information.  

We define $t_\mathrm{E}$ to be the minimum number of leakage events which can cause an \emph{uncorrectable} logical error. 
Following the above discussion, it satisfies $t_\mathrm{E}\leq d_\mathrm{L}$. 
An uncorrectable error occurs when there exists two sets of leakage events which trigger the same erasure checks and may lead to the same syndrome while having different logical outcomes. 
Such an error is uncorrectable even by an ideal decoder. 
We will see that in some settings, the branch-and-bound decoder is necessary to actually correct minimum-weight errors.  

We determine $t_\mathrm{E}$ for each design choice similarly to how we found $d_\mathrm{L}$. 
Specifically, in addition to grouping edges of the detector graph by which leakage events' Pauli envelopes they are in, we group leakage events by which erasure checks they trigger.
We then determine the minimum number of unique erasure checks, using up edges from the Pauli envelopes of up to two leakage events per erasure check, required to traverse the detector graph in the boundary-to-boundary direction. 
The ratio of this minimum number of erasure checks to the ordinary Pauli distance, $\ell$, determines the scaling of $t_\mathrm{E}$ with $\ell$.
As for the leakage fault distance, the boundaries of the detector graph are not considered in this analysis, and therefore we determine only the scaling of $t_\mathrm{E}$ with $\ell$, and neglect any constant terms. 

Results for $t_\mathrm{E}$ are presented in Table~\ref{tab:erasure-t}.

\begin{table}[htb]
    \centering
    \begin{tblr}{
        colspec={@{}l *{4}{c} @{}}, 
        row{2}={m}}
    \toprule
    & \SetCell[c=4]{c}{Surface code circuit} \\
    \cmidrule{2-5}
    & \SetCell[c=2]{c}{all} & & {static and\\moonwalking} & {walking}  \\
    \cmidrule[wd=0.6pt,r=-0.5]{2-3}\cmidrule[wd=0.6pt,l=-0.5,r=-0.5]{4}\cmidrule[wd=0.6pt,l=-0.5]{5}
    \SetCell[r=2]{l,f}{{Leaked-qubit\\effect model}} & \SetCell[c=4]{c}{Erasure check schedule} \\
    \cmidrule{2-5}
                       & 1      & 2                     & 4 or 8               & 4 or 8     \\ \midrule
    Depolarizing       & $\ell$ & $\frac{1}{2}\ell$     & $\frac{1}{2}\ell$   & $\frac{1}{2}\ell$ \\
    Tailored           & $\ell$ & $\ell$                & $\frac{1}{2}\ell$   & $\frac{1}{2}\ell$ \\
    Skip-gate          & $\ell$ & $\ell$                & \hphantom{$^\dag$}$\ell^\dag$         & \hphantom{$^*$}$\frac{2}{3}\ell^*$\\
    \bottomrule
    \end{tblr}
    \caption{
        Scaling of the minimum number of leakage events which can cause an uncorrectable logical error even with erasure information, $t_\mathrm{E}$, for an $\ell\times\ell$ surface code. 
        In all cases except that marked $*$, we find $t_\mathrm{E}$ and $d_\mathrm{L}$ have the same scaling with $\ell$ (see Table~\ref{tab:leakage-d}). 
        For the result marked $\dag$, the branch-and-bound decoder is required to reach the listed {$t_\mathrm{E}\sim\ell$}.
        In all other cases, both decoding strategies can correct all leakage faults with fewer than $t_\mathrm{E}$ leakage events.
        Notice the overall trend that when the leaked-qubit effect is more structured, less-frequent erasure checks suffice to maintain the ideal {$t_\mathrm{E}\sim\ell$}. 
    }
    \label{tab:erasure-t}
\end{table}

Comparing Tables~\ref{tab:leakage-d} and \ref{tab:erasure-t}, we see that for most combinations of leaked-qubit effect, syndrome extraction circuit, and erasure check schedule, the scaling of the minimum number of leakage events that can cause an uncorrectable error, $t_\mathrm{E}$, matches the scaling of the minimum number of leakage events that can cause an undetectable error in the absence of erasure information, $d_\mathrm{L}$. 
That being said, the walking surface code with EC schedule 4 or 8 under the skip-gate leaked-qubit effect, marked (*) in Table~\ref{tab:erasure-t}, exhibits a notable difference. In this case, we find $t_\mathrm{E}\sim\frac{2}{3}\ell$ even though $d_\mathrm{L}\sim\ell$. 
For example, there exist two pairs of leakage events, $\{L_1,L_2\}_A$ and $\{L_1,L_2\}_B$, in a $3\times 3$ circuit which trigger the same erasure checks, yet one pair flips the logical outcome while the other does when they generate the same Pauli syndrome. 
Several such pairs of sets of leakage events exist; see Appendix~\ref{app:t_E} for an example.

The fact that $t_\mathrm{E}<d_\mathrm{L}$ for the walking surface code for the skip-gate leaked-qubit effect and EC schedule 8 was our initial motivation for developing the moonwalking surface code. Indeed, while
the static surface code can maintain $t_\mathrm{E}=d_\mathrm{L}=\ell$ with EC schedules 4 and 8, they both come with an operational overhead: EC schedule 4 requires mid-circuit erasure checks on data qubits during syndrome extraction, and EC schedule 8 requires a leakage-SWAP gate when using the static circuit. 
In contrast, the moonwalking surface code has $t_\mathrm{E}=d_\mathrm{L}=\ell$ under the skip-gate leaked-qubit effect, including for EC schedule 8, which has zero overhead. 
The cause of the differing erasure performance between the walking and moonwalking surface codes is the orientation of their hook errors: the original walking surface code has hook errors parallel to the logical operator of the same type while the moonwalking surface code's hook error is perpendicular to the logical operator. 
The parallel alignment of the walking surface code's hook error is not distance-reducing in the case of stochastic Pauli errors because the gate order reverses every round, preventing these spacetime errors from being tiled across the circuit \cite{gidneyStackexchangeGateOrder2025}. 
However, gate-order alternation cannot fully mitigate the effect of this harmful alignment in the context of infrequent erasure checks. 

To reach $t_\mathrm{E}=d_\mathrm{L}=\ell$, the static and moonwalking surface codes with the skip-gate leaked-qubit effect and EC schedule 4 or 8 require a decoder that incorporates the constraint that a qubit can only leak once. This \emph{leakage disjointness constraint} ensures that an erasure check corresponds to exactly one leakage event. 
Among the decoders presented in Sec.~\ref{sec:decoding}, our branch-and-bound decoder incorporates this constraint, while the marginal decoder does not.
An example of a situation where the errors caused by two leakage events in the $3\times 3$ moonwalking surface code cannot be corrected by the marginal decoder but can be corrected by the branch-and-bound decoder is shown in Appendix~\ref{app:t_E}.

When leakage/erasure is the only type of error present in the system, the LER, $p_\mathrm{L}$, will scale with the physical error rate, $p$, as $p_\mathrm{L}\propto p^{t_\mathrm{E}}$ in the asymptotic regime where $p\rightarrow 0$. 
Outside the asymptotic regime, as long as the physical error rate is well below threshold, the LER scaling exponent is lower-bounded by $t_\mathrm{E}$, i.e., $p_\mathrm{L}\propto p^{\alpha \ell}$ where $\alpha \ell \geq t_\mathrm{E}$.
We verify this in the next section by sampling the LER. 

\subsection{Logical error rate scaling and threshold}\label{sec:LER}

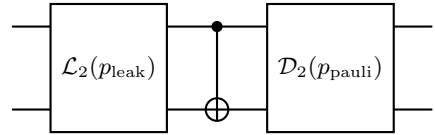
\begin{figure}[htb]
    \centering
    \begin{quantikz}
        & \gate[2]{\mathcal{L}_2(\pleak)} & \ctrl{1} & \gate[2]{\mathcal{D}_2(\pPauli)} & \\
        & & \targ{}  & & 
    \end{quantikz} 
    \caption{
        Sampled noisy circuit error model. Errors are inserted around CX gates.
        Before each CX, there is a probability $\pleak$ for one of two qubits to leak.
        After each CX, we apply the standard two-qubit depolarizing channel with strength $\pPauli$. 
        The erasure bias is $\eta_\mathrm{E}=\pleak/\pPauli$.
    }
    \label{fig:error-model}
\end{figure}

We sample circuits with stochastic leakage before each CX and stochastic Pauli errors after each CX gate; see Fig.~\ref{fig:error-model}. 
The two-qubit leakage channel, $\mathcal{L}_2(\pleak)$, is sampled in two steps:
first, a leakage event is assigned to the gate with probability $\pleak$. Conditional on this event occurring, one of the two qubits is chosen uniformly at random to be the leaked qubit.
In this model, it is not possible for both qubits in a CX to leak at the same time. 
We do not expect this constraint to have a substantial effect, since our LER results presented below are consistent with our conclusions from the previous section, which did not include it. 

For each design choice and code size $\ell\in\{3,5,7,9,11\}$ {(branch-and-bound decoder only used for $\ell\leq 7$)}, we generate the QEC circuit with {$3\ell+1$} syndrome extraction rounds and sample its erasure check, detector, and logical outcomes to determine its logical error rate for a given physical error rate and erasure bias.
The erasure bias is $\eta_\mathrm{E}=\pleak/\pPauli$.
Sampling a circuit has two steps. 
First, we draw one sample from the stochastic leakage processes and directly modify the circuit, adding Pauli noise channels and removing two-qubit gates based on the set of sampled leakage events and the leaked-qubit effect model. The Pauli envelope is not used in this step---only for decoding. 
Each leakage persists until it is removed by an erasure check or three-state measurement, the frequency of which depends on the EC schedule.  
The erasure check outcomes are determined by the set of sampled leakage events and the EC schedule.
The modified circuit is then passed to Stim~\cite{stim} to draw one sample of the detector and logical outcomes. 
We decode based on the erasure check and detector outcomes using the marginal strategy and, for some design choices and $\ell\leq 7$, also the branch-and-bound strategy. 
We declare a logical error if the decoded solution does not match the logical outcome sampled from Stim. 
The probability of logical error after {$3\ell+1$} rounds is normalized by the number of rounds to find the logical error rate (LER) per round, $p_\mathrm{L}$.

\begin{figure}[htb]
    \centering
    \includegraphics{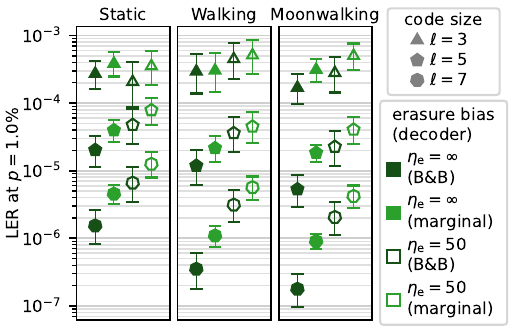}
    \caption{
    Simulated logical error rate (LER) for each surface code circuit at a fixed physical error rate, $p=0.01$, with the skip-gate leaked-qubit effect and erasure check (EC) schedule 8. 
    The $x$ axes are not meaningful. 
    Marker shape indicates code distance and marker filling indicates erasure bias $\eta_\mathrm{e}=\pleak/\pPauli$. 
    Our new moonwalking surface code (SC) circuit significantly outperforms the static (i.e., standard rotated) SC and somewhat outperforms the original walking SC.
    Error bars show a likelihood ratio of 1000.
    }
    \label{fig:pL-vs-circuit}
\end{figure}

\begin{figure*}[!t]
    \centering
    \includegraphics{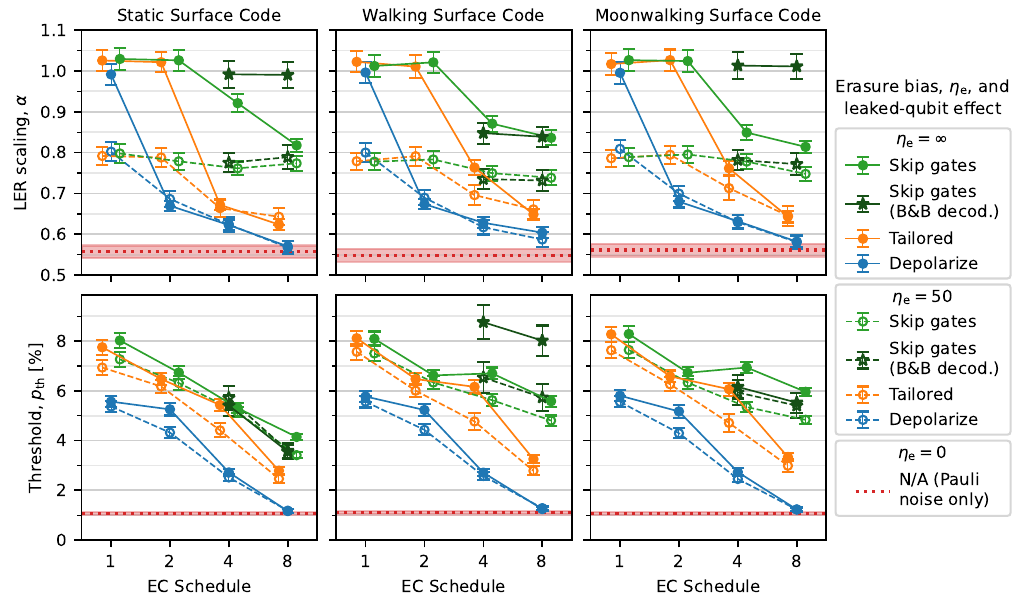}
    \caption{
    Logical error rate (LER) scaling exponent, $\alpha$ (top row), and threshold, $p_\mathrm{th}$ (bottom row), extracted by fitting simulated data to Eq.~\eqref{eq:pL-fit}.
    The different EC schedules as shown on the $x$ axis are categorical, and small horizontal offsets between points for a given schedule are added for legibility. 
    Marker filling and line dashing indicate erasure bias, $\eta_\mathrm{e}=\pleak/\pPauli$. When there is only Pauli noise, EC schedule is irrelevant, so we present that data as a horizontal line. 
    Color indicates leaked-qubit effect. 
    All data is for marginal decoding unless noted otherwise. 
    Notice that branch-and-bound decoding allows one to recover erasure-like scaling, $\alpha\approx 1$, with EC schedules 4 and 8, infinite erasure bias ($\eta_\mathrm{e}$), and skip-gate leakage for the static and moonwalking surface codes (SC), but not for the walking SC, as predicted by the minimum uncorrectable error weights in Table~\ref{tab:erasure-t}. 
    The branch-and-bound decoder does improve the LER for the walking SC (see also Fig.~\ref{fig:pL-vs-circuit}), which manifests here as an improvement in the threshold, $p_\mathrm{th}$, only.
    An improvement in LER scaling is more advantageous than an improvement in threshold as it compounds as physical error rate is decreased and code distance is increased.
    We observe that the moonwalking SC has higher thresholds than the static SC while having the same scaling. 
    Error bars and bands show the standard deviation of fitted parameters using 1000-fold likelihood ratios as the input uncertainty of the data. 
    }
    \label{fig:fit-results}
\end{figure*}

Our simulation results in Fig.~\ref{fig:pL-vs-circuit} show that our moonwalking surface code outperforms both the static and the walking surface codes in the setting we are most interested in, namely, using EC schedule 8, which has no mid-circuit erasure checks, under the skip-gate leaked-qubit effect model.
While our results from Sec.~\ref{sec:min-weights} predict the LER scaling, Fig.~\ref{fig:pL-vs-circuit} shows a single physical error probability, $p$, revealing the contrast between the static and moonwalking surface codes.

To empirically validate the predicted LER scaling, we vary both the physical error rate, $p$, and code size, $\ell$, for each design choice. 
The logical error rate (LER) results are fit to 
\begin{equation}\label{eq:pL-fit}
    p_\mathrm{L}=c (p/p_\mathrm{th})^{\alpha \ell}
\end{equation}
to determine the LER exponent scaling, $\alpha$, and threshold, $p_\mathrm{th}$ for each leaked-qubit effect model, EC schedule, circuit, decoding, and erasure bias. 
To ensure consistency and exclude near-threshold behavior, only points with $p_\mathrm{L}\leq 10^{-3}$ are used for fitting. We present these results in Fig.~\ref{fig:fit-results}. 
Full LER data and overlaid fits are provided in the Supplementary Materials.
We are particularly interested in the LER scaling, $\alpha$, as it varies between $\alpha=1$, which indicates erasure-like QEC performance, and $\alpha=0.5$, indicating Pauli-like performance. 
The extracted LER scaling rates are consistent with the minimum number of leakages required to cause an uncorrectable logical error, $t_\mathrm{E}$, found in Sec.~\ref{sec:min-weights}.
That is, in cases where $t_\mathrm{E}\sim\ell$, we find $\alpha\approx 1$, where $t_\mathrm{E}\sim\frac{2}{3}\ell$, we find $0.67<\alpha<1$, and where $t_\mathrm{E}\sim\frac{1}{2}\ell$, we find $0.5<\alpha<1$. Importantly, these results confirm our prediction that the walking surface code with skip-gate leaked-qubit effect performs worse than the static or moonwalking surface codes for both EC schedules 4 or 8. Though, to fully realize these benefits, one requires a decoder that incorporates the leakage disjointness constraint, such as our branch-and-bound decoder.

\section{Conclusions and outlook}
\label{sec:conclusion}

Our work maps out erasure qubit and circuit design space and shows how the logical error rate scaling and threshold depend on the effect of leaked qubits on other qubits, the schedule of erasure checks, and the erasure bias for three different circuits based on the rotated surface code. This survey will help the QEC community make more informed choices when engineering their erasure qubits.
Importantly, our analytical and numerical results show that one may achieve zero-overhead erasure-like LER scaling without mid-circuit erasure checks, i.e., $p_\mathrm{L}\propto p^\ell$, when our new moonwalking surface code on an $\ell\times\ell$ lattice is augmented with three-state measurement if leaked qubits cause two-qubit gates to be skipped and an appropriate decoder is used.  

Our work motivates several new directions for future research. 
First, while the branch-and-bound decoder proposed in this work recovers the erasure-like scaling for the moonwalking code, this decoder is very slow. Thus, it would be interesting to further investigate the theoretical complexity of the decoding problem with the very structured anticorrelated errors that can arise in the skip-gate leaked-qubit effect model and determine
if fast and accurate decoding is possible. 
Second, in order to be immediately useful to experimentalists, it would be helpful to include two effects we have neglected in
this work: seepage
(the spontaneous return of leaked qubits to the computational
subspace) and incorrect erasure information. 
Finally, we found vastly different performances of the moonwalking and walking codes under skip-gate leaked-qubit effect model and with three-state state readout.
Understanding the fundamental reason behind this discrepancy in a way that enables design of circuits for leakage mitigation and decoders in other QEC codes remains an open direction of research. In particular, we expect error correction circuits based on foliation~\cite{raussendorfFaulttolerantOnewayQuantum2006,raussendorf2007topological,bolt2016foliated,brown2020universal,claes2023tailored} to have a favorable scaling to leakage since the leakage has a much more confined Pauli envelope. 
This is because, as explained in Appendix~\ref{app:pauli-envelope}, the size of a Pauli envelope is determined in part by the number of times a qubit changes from being a control to the target of the CX (and vice versa) between the times it was initialized and reset. Foliation based error correction circuits avoid this problem~\cite{tominprep};
however, the total time cost of foliation is generally larger than conventional syndrome extraction techniques. In a future work we will analyze the conditions under which a scaling advantage can be maintained in presence of these competing effects in foliated codes~\cite{tominprep}. 

\emph{Acknowledgements:}
Thank you to Runjiang Bi, Allen Mi, Kaavya Sahay, Tom Smith, and James Teoh for fruitful scientific discussions. 
Thank you to Jacob Curtis for help editing the manuscript. 

This work was supported by the National Science Foundation (NSF) under Grant No. 2137740 and by the Air Force Office of Scientific Research (AFOSR) under Grant No. FA9550-21-1-0209. 
Any opinions, findings, and conclusions or recommendations expressed in this work are those of the author and do not necessarily reflect the views of the NSF or AFOSR.
CW acknowledges support from the U.S. Department of Energy, Office of Science, National Quantum Information Science Research Centers, Co-Design Center for Quantum Advantage (C2QA) under contract DE-SC0012704. IR acknowledges funding from the Swiss National Science Foundation (Postdoc.Mobility Grant No.~P500-2$\_$235497).  

\bibliography{gf.bib}

@misc{gidneyStackexchangeGateOrder2025,
  title = {Answer to ``{{Is}} There an Algorithm to Determine the Scheduling of {{CNOTs}} for Rotated Surface Codes?''},
  shorttitle = {Answer to ``{{Is}} There an Algorithm to Determine the Scheduling of {{CNOTs}} for Rotated Surface Codes?''},
  author = {Gidney, Craig},
  year = 2025,
  month = aug,
  journal = {Quantum Computing Stack Exchange}
}

@misc{githubSCLE,
  title = {github.com/magzpavz/surface-code-leakage-erasure},
  author = {Pavlovich, Margaret},
  year = 2026,
  month = jul,
  publisher = {GitHub},
  journal = {GitHub repository},
  url = {https://github.com/magzpavz/surface-code-leakage-erasure}
}

@article{wallmanNoiseTailoringScalable2016,
  title = {Noise Tailoring for Scalable Quantum Computation via Randomized Compiling},
  author = {Wallman, Joel J. and Emerson, Joseph},
  year = 2016,
  month = nov,
  journal = {Phys. Rev. A},
  volume = {94},
  number = {5},
  pages = {052325},
  publisher = {American Physical Society},
  doi = {10.1103/PhysRevA.94.052325}
}

@article{claes2023tailored,
	title = {Tailored cluster states with high threshold under biased noise},
	volume = {9},
	copyright = {2023 The Author(s)},
	issn = {2056-6387},
	url = {https://www.nature.com/articles/s41534-023-00677-w},
	doi = {10.1038/s41534-023-00677-w},
	number = {1},
	journal = {npj Quantum Inf.},
	publisher = {Nature Publishing Group},
	author = {Claes, Jahan and Bourassa, J. Eli and Puri, Shruti},
	month = jan,
	year = {2023},
	pages = {9},
}

@unpublished{tominprep,
  author={Smith, Thomas B and others},
  note={In Preparation}
}

@article{bolt2016foliated,
	title = {Foliated {Quantum} {Error}-{Correcting} {Codes}},
	volume = {117},
	url = {https://link.aps.org/doi/10.1103/PhysRevLett.117.070501},
	doi = {10.1103/PhysRevLett.117.070501},
	number = {7},
	journal = {Phys. Rev. Lett.},
	publisher = {American Physical Society},
	author = {Bolt, A. and Duclos-Cianci, G. and Poulin, D. and Stace, T. M.},
	month = aug,
	year = {2016},
	pages = {070501},
}

@article{brown2020universal,
	title = {Universal fault-tolerant measurement-based quantum computation},
	volume = {2},
	url = {https://link.aps.org/doi/10.1103/PhysRevResearch.2.033305},
	doi = {10.1103/PhysRevResearch.2.033305},
	number = {3},
	journal = {Phys. Rev. Res.},
	publisher = {American Physical Society},
	author = {Brown, Benjamin J. and Roberts, Sam},
	month = aug,
	year = {2020},
	pages = {033305},
}

@article{raussendorf2007topological,
	title = {Topological fault-tolerance in cluster state quantum computation},
	volume = {9},
	issn = {1367-2630},
	url = {https://doi.org/10.1088/1367-2630/9/6/199},
	doi = {10.1088/1367-2630/9/6/199},
	number = {6},
	journal = {New J. Phys.},
	author = {Raussendorf, R and Harrington, J and Goyal, K},
	month = jun,
	year = {2007},
	pages = {199},
}

@article{guOptimizingQuantumErrorCorrection2025,
	title = {Optimizing {Quantum} {Error}-{Correction} {Protocols} with {Erasure} {Qubits}},
	volume = {6},
	url = {https://link.aps.org/doi/10.1103/985g-58gd},
	doi = {10.1103/985g-58gd},
	number = {4},
	journal = {PRX Quantum},
	publisher = {American Physical Society},
	author = {Gu, Shouzhen and Vaknin, Yotam and Retzker, Alex and Kubica, Aleksander},
	month = dec,
	year = {2025},
	pages = {040354},
}

@article{neeley2009emulation,
	title = {Emulation of a {Quantum} {Spin} with a {Superconducting} {Phase} {Qudit}},
	volume = {325},
	url = {https://www.science.org/doi/10.1126/science.1173440},
	doi = {10.1126/science.1173440},
	number = {5941},
	journal = {Science},
	publisher = {American Association for the Advancement of Science},
	author = {Neeley, Matthew and Ansmann, Markus and Bialczak, Radoslaw C. and Hofheinz, Max and Lucero, Erik and O'Connell, Aaron D. and Sank, Daniel and Wang, Haohua and Wenner, James and Cleland, Andrew N. and Geller, Michael R. and Martinis, John M.},
	month = aug,
	year = {2009},
	pages = {722--725},
}

@article{bianchetti2010control,
	title = {Control and {Tomography} of a {Three} {Level} {Superconducting} {Artificial} {Atom}},
	volume = {105},
	url = {https://link.aps.org/doi/10.1103/PhysRevLett.105.223601},
	doi = {10.1103/PhysRevLett.105.223601},
	number = {22},
	journal = {Phys. Rev. Lett.},
	publisher = {American Physical Society},
	author = {Bianchetti, R. and Filipp, S. and Baur, M. and Fink, J. M. and Lang, C. and Steffen, L. and Boissonneault, M. and Blais, A. and Wallraff, A.},
	month = nov,
	year = {2010},
	pages = {223601},
}

@article{chen2023transmon,
	title = {Transmon qubit readout fidelity at the threshold for quantum error correction without a quantum-limited amplifier},
	volume = {9},
	copyright = {2023 The Author(s)},
	issn = {2056-6387},
	url = {https://www.nature.com/articles/s41534-023-00689-6},
	doi = {10.1038/s41534-023-00689-6},
	number = {1},
	journal = {npj Quantum Inf.},
	publisher = {Nature Publishing Group},
	author = {Chen, Liangyu and Li, Hang-Xi and Lu, Yong and Warren, Christopher W. and Križan, Christian J. and Kosen, Sandoko and Rommel, Marcus and Ahmed, Shahnawaz and Osman, Amr and Biznárová, Janka and Fadavi Roudsari, Anita and Lienhard, Benjamin and Caputo, Marco and Grigoras, Kestutis and Grönberg, Leif and Govenius, Joonas and Kockum, Anton Frisk and Delsing, Per and Bylander, Jonas and Tancredi, Giovanna},
	month = mar,
	year = {2023},
	pages = {26},
}

@article{wu2019stern,
	title = {Stern–{Gerlach} detection of neutral-atom qubits in a state-dependent optical lattice},
	volume = {15},
	copyright = {2019 The Author(s), under exclusive licence to Springer Nature Limited},
	issn = {1745-2481},
	url = {https://www.nature.com/articles/s41567-019-0478-8},
	doi = {10.1038/s41567-019-0478-8},
	number = {6},
	journal = {Nat. Phys.},
	publisher = {Nature Publishing Group},
	author = {Wu, Tsung-Yao and Kumar, Aishwarya and Giraldo, Felipe and Weiss, David S.},
	month = jun,
	year = {2019},
	pages = {538--542},
}

@article{bluvstein2026fault,
	title = {A fault-tolerant neutral-atom architecture for universal quantum computation},
	volume = {649},
	copyright = {2025 The Author(s)},
	issn = {1476-4687},
	url = {https://www.nature.com/articles/s41586-025-09848-5},
	doi = {10.1038/s41586-025-09848-5},
	number = {8095},
	journal = {Nature},
	publisher = {Nature Publishing Group},
	author = {Bluvstein, Dolev and Geim, Alexandra A. and Li, Sophie H. and Evered, Simon J. and Bonilla Ataides, J. Pablo and Baranes, Gefen and Gu, Andi and Manovitz, Tom and Xu, Muqing and Kalinowski, Marcin and Majidy, Shayan and Kokail, Christian and Maskara, Nishad and Trapp, Elias C. and Stewart, Luke M. and Hollerith, Simon and Zhou, Hengyun and Gullans, Michael J. and Yelin, Susanne F. and Greiner, Markus and Vuletić, Vladan and Cain, Madelyn and Lukin, Mikhail D.},
	month = jan,
	year = {2026},
	pages = {39--46},
}

@article{ghosh2013understanding,
  title = {{Understanding the effects of leakage in superconducting quantum-error-detection circuits}},
  author = {Ghosh, Joydip and Fowler, Austin G. and Martinis, John M. and Geller, Michael R.},
  journal = {Phys. Rev. A},
  volume = {88},
  issue = {6},
  pages = {062329},
  numpages = {7},
  year = {2013},
  month = {Dec},
  publisher = {American Physical Society},
  doi = {10.1103/PhysRevA.88.062329},
  url = {https://link.aps.org/doi/10.1103/PhysRevA.88.062329}
}

@article{brown2018comparing,
  title = {{Comparing Zeeman qubits to hyperfine qubits in the context of the surface code: $^{174}\mathrm{Yb}^{+}$ and $^{171}\mathrm{Yb}^{+}$}},
  author = {Brown, Natalie C. and Brown, Kenneth R.},
  journal = {Phys. Rev. A},
  volume = {97},
  issue = {5},
  pages = {052301},
  numpages = {6},
  year = {2018},
  month = {May},
  publisher = {American Physical Society},
  doi = {10.1103/PhysRevA.97.052301},
  url = {https://link.aps.org/doi/10.1103/PhysRevA.97.052301}
}

@article{brown2019leakage,
  title = {{Leakage mitigation for quantum error correction using a mixed qubit scheme}},
  author = {Brown, Natalie C. and Brown, Kenneth R.},
  journal = {Phys. Rev. A},
  volume = {100},
  issue = {3},
  pages = {032325},
  numpages = {9},
  year = {2019},
  month = {Sep},
  publisher = {American Physical Society},
  doi = {10.1103/PhysRevA.100.032325},
  url = {https://link.aps.org/doi/10.1103/PhysRevA.100.032325}
}

@article{suchara2015leakage, 
    author = {Suchara, Martin and Cross, Andrew W. and Gambetta, Jay M.}, 
    title = {{Leakage suppression in the Toric code}}, 
    year = {2015}, 
    issue_date = {September 2015}, 
    publisher = {Rinton Press, Incorporated}, 
    address = {Paramus, NJ}, 
    volume = {15}, 
    number = {11–12}, 
    issn = {1533-7146}, 
    journal = {Quantum Info. Comput.}, 
    month = {sep}, 
    pages = {997–1016}, 
    numpages = {20}, 
    doi = {10.5555/2871350.2871358},
    url = {https://dl.acm.org/doi/10.5555/2871350.2871358}
}

@article{bennettCapacitiesQuantumErasure1997,
  title = {Capacities of {{Quantum Erasure Channels}}},
  author = {Bennett, Charles H. and DiVincenzo, David P. and Smolin, John A.},
  year = 1997,
  month = apr,
  journal = {Phys. Rev. Lett.},
  volume = {78},
  number = {16},
  pages = {3217--3220},
  publisher = {American Physical Society},
  doi = {10.1103/PhysRevLett.78.3217},
  url = {https://link.aps.org/doi/10.1103/PhysRevLett.78.3217}
}

@article{brownLeakageMitigationQuantum2019,
  title = {Leakage Mitigation for Quantum Error Correction Using a Mixed Qubit Scheme},
  author = {Brown, Natalie C. and Brown, Kenneth R.},
  year = 2019,
  month = sep,
  journal = {Phys. Rev. A},
  volume = {100},
  number = {3},
  pages = {032325},
  publisher = {American Physical Society},
  doi = {10.1103/PhysRevA.100.032325},
  url = {https://link.aps.org/doi/10.1103/PhysRevA.100.032325}
}

@article{changSurfaceCodeImperfect2025,
  title = {Surface {{Code}} with {{Imperfect Erasure Checks}}},
  author = {Chang, Kathleen and Singh, Shraddha and Claes, Jahan and Sahay, Kaavya and Teoh, James and Puri, Shruti},
  year = 2025,
  month = dec,
  journal = {PRX Quantum},
  volume = {6},
  number = {4},
  pages = {040355},
  publisher = {American Physical Society},
  doi = {10.1103/d1v7-nctj},
  url = {https://link.aps.org/doi/10.1103/d1v7-nctj}
}

@article{fowlerCopingQubitLeakage2013,
  title = {Coping with Qubit Leakage in Topological Codes},
  author = {Fowler, Austin G.},
  year = 2013,
  month = oct,
  journal = {Phys. Rev. A},
  volume = {88},
  number = {4},
  pages = {042308},
  publisher = {American Physical Society},
  doi = {10.1103/PhysRevA.88.042308},
  url = {https://link.aps.org/doi/10.1103/PhysRevA.88.042308}
}

@article{fowlerMWPMdecoding2012,
  title = {Proof of {{Finite Surface Code Threshold}} for {{Matching}}},
  author = {Fowler, Austin G.},
  year = 2012,
  month = nov,
  journal = {Phys. Rev. Lett.},
  volume = {109},
  number = {18},
  pages = {180502},
  publisher = {American Physical Society},
  doi = {10.1103/PhysRevLett.109.180502},
  url = {https://link.aps.org/doi/10.1103/PhysRevLett.109.180502}
}

@article{grasslCodesQuantumErasure1997a,
  title = {Codes for the Quantum Erasure Channel},
  author = {Grassl, M. and Beth, {\relax Th}. and Pellizzari, T.},
  year = 1997,
  month = jul,
  journal = {Phys. Rev. A},
  volume = {56},
  number = {1},
  pages = {33--38},
  publisher = {American Physical Society},
  doi = {10.1103/PhysRevA.56.33},
  url = {https://link.aps.org/doi/10.1103/PhysRevA.56.33}
}

@article{guFaulttolerantQuantumArchitectures2025,
  title = {Fault-Tolerant Quantum Architectures Based on Erasure Qubits},
  author = {Gu, Shouzhen and Retzker, Alex and Kubica, Aleksander},
  year = 2025,
  month = mar,
  journal = {Phys. Rev. Res.},
  volume = {7},
  number = {1},
  pages = {013249},
  publisher = {American Physical Society},
  doi = {10.1103/PhysRevResearch.7.013249},
  url = {https://link.aps.org/doi/10.1103/PhysRevResearch.7.013249}
}

@article{kangQuantumErrorCorrection2023,
  title = {Quantum {{Error Correction}} with {{Metastable States}} of {{Trapped Ions Using Erasure Conversion}}},
  author = {Kang, Mingyu and Campbell, Wesley C. and Brown, Kenneth R.},
  year = 2023,
  month = jun,
  journal = {PRX Quantum},
  volume = {4},
  number = {2},
  pages = {020358},
  publisher = {American Physical Society},
  doi = {10.1103/PRXQuantum.4.020358},
  url = {https://link.aps.org/doi/10.1103/PRXQuantum.4.020358}
}

@article{KLM2001,
  title = {A Scheme for Efficient Quantum Computation with Linear Optics},
  author = {Knill, E. and Laflamme, R. and Milburn, G. J.},
  year = 2001,
  month = jan,
  journal = {Nature},
  volume = {409},
  number = {6816},
  pages = {46--52},
  publisher = {Nature Publishing Group},
  issn = {1476-4687},
  doi = {10.1038/35051009},
  copyright = {2001 Macmillan Magazines Ltd.},
  langid = {english}
}

@article{koottandavidaErasureDetectionDualRail2024,
  title = {Erasure {{Detection}} of a {{Dual-Rail Qubit Encoded}} in a {{Double-Post Superconducting Cavity}}},
  author = {Koottandavida, Akshay and Tsioutsios, Ioannis and Kargioti, Aikaterini and Smith, Cassady R. and Joshi, Vidul R. and Dai, Wei and Teoh, James D. and Curtis, Jacob C. and Frunzio, Luigi and Schoelkopf, Robert J. and Devoret, Michel H.},
  year = 2024,
  month = may,
  journal = {Phys. Rev. Lett.},
  volume = {132},
  number = {18},
  pages = {180601},
  publisher = {American Physical Society},
  doi = {10.1103/PhysRevLett.132.180601},
  url = {https://link.aps.org/doi/10.1103/PhysRevLett.132.180601}
}

@article{kubicaErasureQubitsOvercoming2023a,
  title = {Erasure {{Qubits}}: {{Overcoming}} the $T_1$ {{Limit}} in {{S    uperconducting Circuits}}},
  shorttitle = {Erasure {{Qubits}}},
  author = {Kubica, Aleksander and Haim, Arbel and Vaknin, Yotam and Levine, Harry and Brand{\~a}o, Fernando and Retzker, Alex},
  year = 2023,
  month = nov,
  journal = {Phys. Rev. X},
  volume = {13},
  number = {4},
  pages = {041022},
  publisher = {American Physical Society},
  doi = {10.1103/PhysRevX.13.041022},
  url = {https://link.aps.org/doi/10.1103/PhysRevX.13.041022}
}

@article{levineDemonstratingLongCoherenceDualRail2024,
  title = {Demonstrating a {{Long-Coherence Dual-Rail Erasure Qubit Using Tunable Transmons}}},
  author = {Levine, H. and Haim, A. and Hung, J. S. C. and Alidoust, N. and Kalaee, M. and DeLorenzo, L. and Wollack, E. A. and {Arrangoiz-Arriola}, P. and Khalajhedayati, A. and Sanil, R. and Moradinejad, H. and Vaknin, Y. and Kubica, A. and Hover, D. and Aghaeimeibodi, S. and Alcid, J. A. and Baek, C. and Barnett, J. and Bawdekar, K. and Bienias, P. and Carson, H. A. and Chen, C. and Chen, L. and Chinkezian, H. and Chisholm, E. M. and Clifford, A. and Cosmic, R. and Crisosto, N. and Dalzell, A. M. and Davis, E. and D'Ewart, J. M. and Diez, S. and D'Souza, N. and Dumitrescu, P. T. and Elkhouly, E. and Fang, M. T. and Fang, Y. and Flammia, S. and Fling, M. J. and Garcia, G. and Gharzai, M. K. and Gorshkov, A. V. and Gray, M. J. and Grimberg, S. and Grimsmo, A. L. and Hann, C. T. and He, Y. and Heidel, S. and Howell, S. and Hunt, M. and Iverson, J. and Jarrige, I. and Jiang, L. and Jones, W. M. and Karabalin, R. and Karalekas, P. J. and Keller, A. J. and Lasi, D. and Lee, M. and Ly, V. and MacCabe, G. and Mahuli, N. and Marcaud, G. and Matheny, M. H. and McArdle, S. and McCabe, G. and Merton, G. and Miles, C. and Milsted, A. and Mishra, A. and Moncelsi, L. and Naghiloo, M. and Noh, K. and Oblepias, E. and Ortuno, G. and Owens, J. C. and Pagdilao, J. and Panduro, A. and Paquette, J.-P. and Patel, R. N. and Peairs, G. and Perello, D. J. and Peterson, E. C. and Ponte, S. and Putterman, H. and Refael, G. and Reinhold, P. and Resnick, R. and Reyna, O. A. and Rodriguez, R. and Rose, J. and Rubin, A. H. and Runyan, M. and Ryan, C. A. and Sahmoud, A. and Scaffidi, T. and Shah, B. and Siavoshi, S. and Sivarajah, P. and Skogland, T. and Su, C.-J. and Swenson, L. J. and Sylvia, J. and Teo, S. M. and Tomada, A. and Torlai, G. and Wistrom, M. and Zhang, K. and Zuk, I. and Clerk, A. A. and Brand{\~a}o, F. G. S. L. and Retzker, A. and Painter, O.},
  year = 2024,
  month = mar,
  journal = {Phys. Rev. X},
  volume = {14},
  number = {1},
  pages = {011051},
  publisher = {American Physical Society},
  doi = {10.1103/PhysRevX.14.011051},
  url = {https://link.aps.org/doi/10.1103/PhysRevX.14.011051}
}

@misc{liuAchievingOptimalDistanceAtomLoss2026,
  title={Achieving Optimal-Distance Atom-Loss Correction via Pauli Envelope}, 
  author={Pengyu Liu and Shi Jie Samuel Tan and Eric Huang and Umut A. Acar and Hengyun Zhou and Chen Zhao},
  year={2026},
  eprint={2603.04156},
  archivePrefix={arXiv},
  primaryClass={quant-ph},
  url={https://arxiv.org/abs/2603.04156}, 
  doi = {10.48550/arXiv.2603.04156},
}

@misc{liuHardwareEfficientErasureQubits2026,
  title={Hardware-Efficient Erasure Qubits With Superconducting Transmon Qutrits}, 
  author={Bao-Jie Liu and Ying-Ying Wang and Yu-Xin Wang and Manthan Badbaria and Shruti Puri and Chen Wang},
  year={2026},
  eprint={2604.08672},
  archivePrefix={arXiv},
  primaryClass={quant-ph},
  url={https://arxiv.org/abs/2604.08672}, 
  doi={10.48550/arXiv.2604.08672}
}

@article{maHighfidelityGatesMidcircuit2023,
  title = {High-Fidelity Gates and Mid-Circuit Erasure Conversion in an Atomic Qubit},
  author = {Ma, Shuo and Liu, Genyue and Peng, Pai and Zhang, Bichen and Jandura, Sven and Claes, Jahan and Burgers, Alex P. and Pupillo, Guido and Puri, Shruti and Thompson, Jeff D.},
  year = 2023,
  month = oct,
  journal = {Nature},
  volume = {622},
  number = {7982},
  pages = {279--284},
  publisher = {Nature Publishing Group},
  issn = {1476-4687},
  doi = {10.1038/s41586-023-06438-1},
  copyright = {2023 The Author(s), under exclusive licence to Springer Nature Limited},
}

@article{mcewenRelaxingHardwareRequirements2023,
  title = {Relaxing {H}ardware {R}equirements for {S}urface {C}ode {C}ircuits using {T}ime-dynamics},
  author = {McEwen, Matt and Bacon, Dave and Gidney, Craig},
  journal = {{Quantum}},
  issn = {2521-327X},
  publisher = {{Verein zur F{\"{o}}rderung des Open Access Publizierens in den Quantenwissenschaften}},
  volume = {7},
  pages = {1172},
  month = nov,
  year = {2023},
  doi = {10.22331/q-2023-11-07-1172},
  url = {https://doi.org/10.22331/q-2023-11-07-1172},
}

@misc{mehtaBiaspreservingErrordetectableEntangling2025a,
  title={Bias-preserving and error-detectable entangling operations in a superconducting dual-rail system}, 
  author={Nitish Mehta and James D. Teoh and Taewan Noh and Ankur Agrawal and Amos Anderson and Beau Birdsall and Avadh Brahmbhatt and Winfred Byrd and Marc Cacioppo and Anthony Cabrera and Leo Carroll and Jonathan Chen and Tzu-Chiao Chien and Richard Chamberlain and Jacob C. Curtis and Doreen Danso and Sanjana Renganatha Desigan and Francesco D'Acounto and Bassel Heiba Elfeky and S. M. Farzaneh and Chase Foley and Benjamin Gudlewski and Hannah Hastings and Robert Johnson and Nishaad Khedkar and Trevor Keen and Anup Kumar and Cihan Kurter and Kamila Krawczuk and Eric Langstengel and Richard D. Li and Gangqiang Liu and Hanyi Lu and Pinlei Lu and Luke Mastalli-Kelly and Adam Maines and Michael Maxwell and Heather McCarrick and Mona Mirzaei and Anirudh Narla and Omar Rashad and Erik Reikes and Mizanur Rahman and Rurik Primiani and Michael Schwaller and Ali Sabbah and Tali Shemma and Ruby A. Shi and Sitakanta Satapathy and Dean Stolpe and Jonathan Strenczewilk and Doug Szperka and Iu-Wei Sze and David Sweeney and Preetham Tikkireddi and Chin-Lun Tsung and Daren Vet Sam and Daniel K. Weiss and Zhibo Yang and Liuqi Yu and Teng Zhang and Olivier Boireau and Stephen Horton and Sean Weinberg and Jose Aumentado and Bryan Cord and Chan U Lei and Joseph O. Yuan and Shantanu O. Mundhada and Kevin S. Chou and S. Harvey Moseleley, Jr. and Robert J. Schoelkopf},
  year={2025},
  eprint={2503.10935},
  archivePrefix={arXiv},
  primaryClass={quant-ph},
  url={https://arxiv.org/abs/2503.10935}, 
  doi = {10.48550/arXiv.2503.10935}
}

@article{pymatching,
  title = {Sparse {{Blossom}}: Correcting a Million Errors per Core Second with Minimum-Weight Matching},
  shorttitle = {Sparse {{Blossom}}},
  author = {Higgott, Oscar and Gidney, Craig},
  year = 2025,
  month = jan,
  journal = {Quantum},
  volume = {9},
  pages = {1600},
  publisher = {Verein zur F\"orderung des Open Access Publizierens in den Quantenwissenschaften},
  doi = {10.22331/q-2025-01-20-1600},
  langid = {british}
}

@article{quinnHighfidelityEntanglementMetastable2026,
  title = {High-Fidelity Entanglement of Metastable Trapped-Ion Qubits with Integrated Erasure Conversion},
  author = {Quinn, A. and Gregory, G. J. and Moore, I. D. and Brudney, S. and Metzner, J. and Ritchie, E. R. and O'Reilly, J. and Wineland, D. J. and Allcock, D. T. C.},
  year = 2026,
  month = apr,
  journal = {Phys. Rev. A},
  volume = {113},
  number = {4},
  pages = {L040601},
  publisher = {American Physical Society},
  doi = {10.1103/p3cy-8yjk}
}

@article{raussendorfFaulttolerantOnewayQuantum2006,
  title = {A Fault-Tolerant One-Way Quantum Computer},
  author = {Raussendorf, R. and Harrington, J. and Goyal, K.},
  year = 2006,
  month = sep,
  journal = {Annals of Physics},
  volume = {321},
  number = {9},
  pages = {2242--2270},
  issn = {0003-4916},
  doi = {10.1016/j.aop.2006.01.012}
}

@article{schollErasureConversionHighfidelity2023,
  title = {Erasure Conversion in a High-Fidelity {{Rydberg}} Quantum Simulator},
  author = {Scholl, Pascal and Shaw, Adam L. and Tsai, Richard Bing-Shiun and Finkelstein, Ran and Choi, Joonhee and Endres, Manuel},
  year = 2023,
  month = oct,
  journal = {Nature},
  volume = {622},
  number = {7982},
  pages = {273--278},
  publisher = {Nature Publishing Group},
  issn = {1476-4687},
  doi = {10.1038/s41586-023-06516-4},
  copyright = {2023 The Author(s)},
  langid = {english}
}

@article{stim,
  title = {Stim: A Fast Stabilizer Circuit Simulator},
  shorttitle = {Stim},
  author = {Gidney, Craig},
  year = 2021,
  month = jul,
  journal = {Quantum},
  volume = {5},
  pages = {497},
  publisher = {Verein zur F\"orderung des Open Access Publizierens in den Quantenwissenschaften},
  doi = {10.22331/q-2021-07-06-497},
  langid = {british}
}

@article{teohDualrailEncodingSuperconducting2023,
  title = {Dual-Rail Encoding with Superconducting Cavities},
  author = {Teoh, James D. and Winkel, Patrick and Babla, Harshvardhan K. and Chapman, Benjamin J. and Claes, Jahan and {de Graaf}, Stijn J. and Garmon, John W. O. and Kalfus, William D. and Lu, Yao and Maiti, Aniket and Sahay, Kaavya and Thakur, Neel and Tsunoda, Takahiro and Xue, Sophia H. and Frunzio, Luigi and Girvin, Steven M. and Puri, Shruti and Schoelkopf, Robert J.},
  year = 2023,
  month = oct,
  journal = {Proc. Natl. Acad. Sci. U. S. A.},
  volume = {120},
  number = {41},
  pages = {e2221736120},
  publisher = {Proceedings of the National Academy of Sciences},
  doi = {10.1073/pnas.2221736120},
  url = {https://doi.org/10.1073/pnas.2221736120}
}

@article{wuErasureConversionFaulttolerant2022,
  title = {Erasure Conversion for Fault-Tolerant Quantum Computing in Alkaline Earth {{Rydberg}} Atom Arrays},
  author = {Wu, Yue and Kolkowitz, Shimon and Puri, Shruti and Thompson, Jeff D.},
  year = 2022,
  month = aug,
  journal = {Nat Commun},
  volume = {13},
  number = {1},
  pages = {4657},
  publisher = {Nature Publishing Group},
  issn = {2041-1723},
  doi = {10.1038/s41467-022-32094-6},
  copyright = {2022 The Author(s)},
  langid = {english}
}

@article{yuTamingRydbergDecay2026a,
  title = {Taming {{Rydberg Decay}} with {{Measurement-Based Quantum Computation}}},
  author = {Yu, Cheng-Cheng and Chen, Zi-Han and Deng, Yu-Hao and Lu, Chao-Yang and Chen, Ming-Cheng and Pan, Jian-Wei},
  year = 2026,
  month = apr,
  journal = {Phys. Rev. Lett.},
  volume = {136},
  number = {16},
  pages = {160601},
  publisher = {American Physical Society},
  doi = {10.1103/msbj-fxw7}
}

\appendix

\section{Construction of dynamic syndrome extraction circuits}\label{app:dynamic-circuit-construction}

\begin{figure}[htb]
    \centering
    \begin{quantikz}[column sep=4pt, row sep={16pt,between origins}, align equals at=1.5]
        & \ctrl{1} &[2pt] & \\
        & \targ{}  &  \mz & 
    \end{quantikz}
    $\!=\!$
    \begin{quantikz}[column sep=6pt, row sep={16pt,between origins}, align equals at=1.5]
        & \ctrl{1} &[4pt] \ctrl{1}\gategroup[2,steps=3,style={dashed,inner sep=1pt}]{SWAP} & \targ{}   & \ctrl{1} &[6pt] \swap{1} &[2pt] & \\
        & \targ{}  & \targ{}                                                               & \ctrl{-1} & \targ{}  &      \targX{} &  \mz &
    \end{quantikz}
    $\!=\!$
    \begin{quantikz}[column sep=6pt, row sep={16pt,between origins}, align equals at=1.5]
        & \targ{}   & \ctrl{1} & \mz &[2pt] \swap{1} & \\
        & \ctrl{-1} & \targ{}  &     &      \targX{} &
    \end{quantikz}
    $\!=\!$
    \begin{quantikz}[column sep=6pt, row sep={16pt,between origins}, align equals at=1.5]
        & \targ{}   & \mz\wire[d]{c} &[2pt] \swap{1} & \\
        & \ctrl{-1} &    \targ{}     &      \targX{} &
    \end{quantikz}
    \caption{
    The circuit equivalences and transformations to construct the walking surface code circuit~\cite{mcewenRelaxingHardwareRequirements2023}. Compare to Fig.~\ref{fig:moonwalking-construction}. 
    A pair of SWAP operations is inserted between the final CX gate and the measurement. 
    One SWAP is commuted through the measurement operation to the end of the syndrome extraction circuit. 
    In the context of the whole code, this SWAP is meaningless. Its effect can be recovered by simultaneously shifting all the data qubits along a diagonal by half a plaquette in software so that they appear to be placed on the sublattice formed by ancilla qubits for the next round.
    The other SWAP is decomposed into CX operations which are combined with the existing CX gate and the measurement operation, which has the cumulative effect of flipping the original CX gate and adding a classically-controlled $X$ gate which can be handled in software.
    }
    \label{fig:walking-construction}
\end{figure}
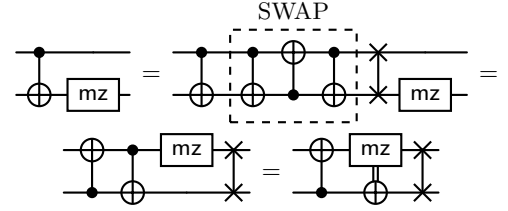

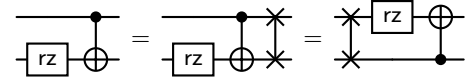
\begin{figure}[htb]
    \centering
    \begin{quantikz}[column sep=4pt, row sep={16pt,between origins}, align equals at=1.5]
        &    &[2pt] \ctrl{1}& \\
        & \rz&      \targ{} & 
    \end{quantikz}
    $\!=\!$
    \begin{quantikz}[column sep=4pt, row sep={16pt,between origins}, align equals at=1.5]
        &    &[2pt] \ctrl{1} &[4pt] \swap{1} &[2pt] \\
        & \rz&      \targ{}  &      \targX{} &
    \end{quantikz}
    $\!=\!$
    \begin{quantikz}[column sep=4pt, row sep={16pt,between origins}, align equals at=1.5]
        &[2pt] \swap{1} &[4pt] \rz&[2pt] \targ{} & \\
        &      \targX{} &    & \ctrl{-1}& 
    \end{quantikz}
    \caption{
    The moonwalking surface code circuit may also be constructed based on mid-SWAP circuit of Ref.~\cite{liuAchievingOptimalDistanceAtomLoss2026}, rather than using the equivalences shown in Fig.~\ref{fig:moonwalking-construction}. 
    The first equivalence here holds because the two qubits are entangled as $\alpha\ket{00}+\beta\ket{11}$ after the reset and CX, and therefore may be freely swapped~\cite{liuAchievingOptimalDistanceAtomLoss2026}. 
    The second equivalence comes from commuting the SWAP to the left. 
    }
    \label{fig:mid-SWAP}
\end{figure}

See Fig.~\ref{fig:walking-construction} for the construction of the walking surface code and Fig.~\ref{fig:mid-SWAP} for an alternate construction of the moonwalking surface code from that presented in Fig.~\ref{fig:moonwalking-construction}.

\section{Decoding}\label{app:decoding}

The task of most likely error (MLE) decoding is to find which physical error or fault, $F$, is most likely to have happened in a QEC circuit given the observed syndrome, $\bar{S}$. The syndrome is the pattern of detectors that were triggered. Each detector is a product of measurement outcomes, e.g., of a stabilizer measurement outcome in two consecutive rounds. Note that a fault may consist of many individual errors across the entire circuit volume. 
That is, we wish to find 
\begin{equation}\label{eq:most-likely-error}
    F^\star = \argmax_{F\in\Omega} p(F)\quad \text{s.t.}\quad S(F) = \bar{S},
\end{equation}
where $\Omega$ is the set of all possible faults, $p(F)$ is the probability of fault $F$, and $S(F)$ is the syndrome that fault $F$ gives rise to.
Decoding is successful if $F^\star$ is truly the fault that happened or equivalent to it in terms of its effect on the logical operator(s). 

If all possible errors in the circuit are Pauli operators that are applied independently and stochastically, then this problem may be cast to a hypergraph-matching problem in general and approximately to a graph-matching problem in the case of the surface code and its variants considered in this work~\cite{fowlerMWPMdecoding2012}; see Appendix~\ref{app:dems-graphs}. By casting the problem to graph-matching, we gain access to fast algorithms \cite{pymatching}.

There are two equivalent perspectives one may take on erasure information and its relationship to decoding: 
one can say it provides information about the fault probabilities, $p(F)$, or one may view the syndrome as including the erasure information. 
In either case, erasures can enable one to decode more accurately but may break the equivalence to a (hyper)graph-matching problem (see Appendix~\ref{app:dems-graphs}). 
In Appendix~\ref{app:pauli-envelope}, we describe how we can use the machinery developed for dealing with Pauli errors in the case of leakages. 
Then, we outline the two strategies we deploy for decoding with erasures: in Appendix~\ref{app:marginal-decoding}, we describe the \emph{marginal} strategy, which returns us to a graph problem at the cost of accuracy in some cases; and in Appendix~\ref{app:mle-decoding}, we describe the \emph{quasi-MLE} strategy, which solves the most-likely fault problem almost exactly but requires more computational resources.

Both strategies make two approximations: 
(1) Detector error model (DEM) hypergraphs are decomposed into independent $X$ and $Z$ graphs (see Appendix~\ref{app:dems-graphs}) and (2) each leakage location that could trigger a given erasure check is equally likely \footnote{
    The probability that the leakage occurred at the $j$th (where $j=1$ is the earliest) possible leakage location for erasure check $i$ is $P_{i,j}=\frac{(1-p)^{j-1}p}{1-(1-p)^{N_i}}\approx \frac{1-(j-1)p}{N_i}$, where $p$ is the \textit{a priori} probability of leakage at any given location and $N_i$ is the total number of possible leakage locations that could trigger erasure check $i$. The deviation of this from a flat $1/{N_i}$ distribution is negligible as $p\lesssim 10^{-2}$ and ${N_i}\leq 8$.
}. Neither approximation is distance-reducing in this setting. 
The marginal decoding strategy additionally approximates disjoint errors as independent which can reduce effective distance. 
These disjoint errors arise because a qubit may only leak once between erasure checks in our model. 

\subsection{Detector error models and graphs}\label{app:dems-graphs}

Both decoding strategies we employ in this work are based on calculating and manipulating detector error models (DEMs). A DEM is simply the set of errors that may occur in a circuit. For each error, we have its probability, which detectors it flips, and whether it flips the logical outcome(s). 
A matching hypergraph may be constructed from a DEM if the errors are assumed to be independent, which is the conventional assumption. The need for this assumption is explained below.
In this hypergraph, the detectors comprise the vertices and the errors comprise the hyperedges, which are assigned weights, $w$, according to the errors' probabilities, $p$: $w(p)\coloneq\log\left( (1-p)/p \right)$. 
If two errors have the same effect (their hyperedges are parallel), they are merged and the new probability is $(p_1,p_2)\rightarrow p_1(1-p_2)+(1-p_1)p_2$. 
The hypergraph is a simple graph if each error triggers exactly one or two detectors. In the case of the surface code, one may decompose the original hypergraph into independent $X$ and $Z$ graphs by approximating $Y$ errors as coincident $X$ and $Z$ errors.  
Calculating a circuit's DEM and converting it into a graph converts the decoding problem into a minimum-weight perfect matching problem, which can be solved efficiently. 

Decoding is equivalent to hypergraph matching only if the errors are independent because this lets us relate the probability of a joint error to the sum of the corresponding edge weights.
Consider a DEM with the set of errors $E=\{e_1,\dots,e_N\}$ which individually have probabilities $p[e_i]$. 
Given a subset of errors, $\bar{E}\subset E$, 
the probability of fault $\bar{F}=\prod_{e\in \bar{E}}e$ is 
\begin{equation*}
    p(\bar{F})=\big(\prod_{e\in \bar{E}}p[e]\big)\big(\prod_{e\in E\setminus\bar{E}}(1-p[e])\big)
\end{equation*}
if and only if the errors in $E$ are independent \footnote{
    This expression for the probability of a fault also assumes that there is only one decomposition for the fault. 
    Since the goal here is simply to demonstrate the necessity of the statistical independence of errors, this does not affect the conclusion.
}. 
The weight of the fault on the decoding hypergraph is the sum of the weights of the errors in $\bar{E}$:
\begin{equation*}
    {w[\bar{F}]= \sum_{e\in \bar{E}}w[e] = \sum_{e\in \bar{E}} \log\left( (1-p[e])/p[e] \right) }.
\end{equation*}
Thus, we have $w[\bar{F}]=\log(A/p[\bar{F}])$ with scaling factor ${A=\left(\prod_{e\in E}(1-p[e])\right)}$, 
showing that the total weight of a fault on the decoding hypergraph is monotonically related to its probability. 
Because the task of decoding is to find the most likely fault for a given syndrome, $F^\star$, Eq.~\eqref{eq:most-likely-error}, we can monotonically transform the probabilities without affecting the result---the fault with the highest probability also has the lowest weight. 
Therefore, decoding is equivalent to hypergraph matching when errors are independent.

As discussed in Sec.~\ref{sec:decoding}, 
the challenge of decoding with infrequent erasure checks is precisely because the errors are not independent. 
When an erasure check is triggered, we know that the qubit leaked a some point since the previous erasure check. At which time step the qubit leaked is a random variable. 
Crucially, a qubit may only leak once (including seepage, which is neglected in this work, would make the leakage locations no longer strictly disjoint; however, leakage locations would remain anti-correlated).
Therefore, the leakage time is described by a disjoint probability distribution, not by independent probabilities of leakage at each time step. 

We note again that the ability to formulate the decoding problem as a hypergraph-matching problem is separate from the ability to decompose the decoding hypergraph into simple graphs. The first requires errors to not be anti-correlated. The second requires single errors to trigger no more than two detectors and for errors to not be correlated. 

\subsubsection{Notation}

We now show the types of manipulations we perform on DEMs in Appendices~\ref{app:marginal-decoding}--\ref{app:mle-decoding} and explain our notation. 
There are three operations we perform on DEMs: scalar multiplication and two types of addition/combination. The two types of DEM addition are independent and disjoint. 
We are only interested in adding DEMs for the same circuit but different error types and locations, e.g., stochastic Pauli errors or leakages at specific circuit locations. 
Independent DEM addition is used to combine DEMs from different erasure checks. 
We will use scalar multiplication and disjoint DEM addition to combine DEMs from different leakage events associated with the same erasure check. 
We explain how these DEMs are determined after this section.  

Going forward, 
we assume that DEMs have been decomposed into graphs; a DEM can be represented by the graph $G=(V,E,w)$, with vertices (i.e., detectors) $V$, edges (i.e., errors or products of errors) $E$, and weights $w[e]$. The weights are determined by the error probabilities: $w[e]=w\left( p[e] \right)$, where 
\begin{equation}\label{eq:w(p)}
    w(p)\coloneq\log\left( (1-p)/p \right)
\end{equation}
and $p[e]$ is the probability of edge $e\in E$. When a graph does not include a given edge, we may equivalently say that the edge has 0 probability and infinite weight: $e\notin E\rightarrow p[e]=0,\,w[e]=\infty$.

Scalar multiplication rescales all probabilities in a DEM. 
Given a DEM graph $G=(V,E,w)$, we scale it by $0\leq a\leq 1$ as follows:
\begin{align}\label{eq:dem-scalar-mult}
    G' &= a G = (V, E, w') \\
    w'[e] &= w(p'[e]) \text{ where } p'[e]=ap[e]. \nonumber
\end{align}
This operation always increases edge weights or leaves them unchanged: $p'[e]\leq p[e]\rightarrow w'[e]\geq w[e]$. 

The independent sum of DEMs $G_1=(V,E_1,w_1)$ and $G_2=(V,E_2,w_2)$ is
\begin{align}\label{eq:dem-comb-indep}
    G_1+G_2 &= (V, E_1 \cup E_2, w_\mathrm{ind}), \\ 
    w_\mathrm{ind}[e] &= w\left(p_\mathrm{ind}[e]\right), \nonumber \\ 
    p_\mathrm{ind}[e] &= p_1[e](1-p_2[e])+(1-p_1[e])p_2[e],\nonumber
\end{align}
where we use sum notation due to the similarity to sums of random variables, not a conventional graph sum. 
Notice that the input graphs $G_1$ and $G_2$ have the same set of vertices $V$ because they reference the same underlying error-free circuit. 
Assuming the input probabilities satisfy $p_1, p_2\leq 0.5$ (which is always true in this work), then independent addition always decreases edge weights or leaves them unchanged: $p_\mathrm{ind}\geq p_1,p_2\rightarrow w_\mathrm{ind}\leq w_1, w_2$.
We add DEMs independently when their error probabilities are truly independent, i.e., the DEMs describe independent stochastic processes. 
Then, $p_\mathrm{ind}[e]$ is the probability of the event $e$ happening exactly once. 

The disjoint sum of DEMs $G_1=(V,E_1,w_1)$ and $G_2=(V,E_2,w_2)$ is
\begin{align}\label{eq:dem-comb-disj}
    G_1 \oplus G_2 &= (V, E_1 \cup E_2, w_\mathrm{disj}),  \\
    w_\mathrm{disj}[e] &= w\left(p_\mathrm{disj}[e]\right), \nonumber \\ 
    p_\mathrm{disj}[e] &= p_1[e]+p_2[e]. \nonumber
\end{align}
Disjoint addition always decreases edge weights or leaves them unchanged: $p_\mathrm{disj}\geq p_1,p_2\rightarrow w_\mathrm{disj}\leq w_1, w_2$.
Of course, this operation is only sensible if $p_1+p_2\leq 1$, which is satisfied in our actual use case. 
We use disjoint sums when only one DEM describes reality, but we don't know which one. 
Given probabilities $P_1$ and $P_2$ of DEM $G_1$ or $G_2$ being correct, respectively, the disjoint-sum DEM $G_3=(P_1G_1)\oplus(P_2G_2)$ has the proper marginal probabilities for each event occurring. 
Specifically, the probability of an event $e$ in the sum DEM, $G_3$, is 
\begin{align*}
    p_3[e] = & P_1p_1[e]+P_2p_2[e] \\
     =&P(\text{DEM 1 is true})P(\text{event }e\,|\,\text{DEM 1 is true})\\
       &+P(\text{DEM 2 is true})P(\text{event }e\,|\,\text{DEM 2 is true}).
\end{align*} 
Because $P_1+P_2\leq 1$, it is guaranteed that the sum DEM has probabilities no greater than the larger of the input probabilities. 

\subsection{Pauli envelopes}\label{app:pauli-envelope}

The impact of a particular leakage event can be bounded by a Pauli channel called its \emph{Pauli envelope} \cite{liuAchievingOptimalDistanceAtomLoss2026}. 
The Pauli envelope framework was introduced in \cite{liuAchievingOptimalDistanceAtomLoss2026}. We expand its usage to different leakage effects and provide an intuitive derivation of the Pauli envelope for skip-gate leakage. 
The envelope depends on the circuit location of the leakage event, the effect of leakage on other qubits (Sec.~\ref{sec:leaked-qubit-effects}), and the erasure check schedule (Sec.~\ref{sec:ec-sched}). 

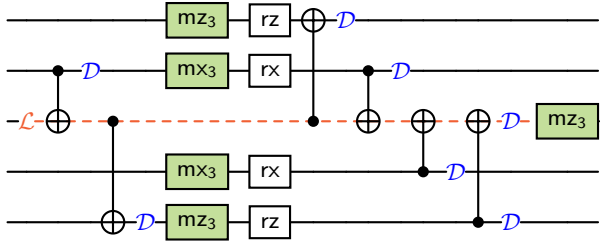
\begin{figure}[htb]
    \centering
    \begin{quantikz}[row sep={6pt}, column sep={4pt}]
    &          &                       &       & [-4pt]  &       & \mzEC& [4pt] \rz& \targ{}  & \depol& [-4pt]  &       & [-4pt]   &       & [-4pt]   &                                               & [2pt]& [-4pt]             \\
    &          & \ctrl{1}              & \depol&         &       & \mxEC& \rx      &          &       & \ctrl{1}& \depol&          &       &          &                                               &      & \\
    & \leak{14}& \targ{}\setwiretype{n}&       & \ctrl{2}&       &      &          & \ctrl{-2}&       & \targ{} &       & \targ{}  &       & \targ{}  & \depol\wire[r][1][style={RedOrange,dashed}]{q}& \mzEC& \setwiretype{q}   \\
    &          &                       &       &         &       & \mxEC& \rx      &          &       &         &       & \ctrl{-1}& \depol&          &                                               &      & \\
    &          &                       &       & \targ{} & \depol& \mzEC& \rz      &          &       &         &       &          &       & \ctrl{-2}& \depol                                        &      & 
    \end{quantikz}
    \caption{Example of a leakage event's Pauli envelope for depolarizing leakage and EC schedule 8 on the moonwalking surface code. 
    The green-filled $\mathsf{mz_3}$ ($\mathsf{mx_3}$) is the three-state measurement which returns erasure information and $Z$-basis ($X$-basis) logical information. 
    The $\mathsf{r}$ ($\mathsf{rx}$) gate resets the qubit to $\ket{0}$ ($\ket{+}$). 
    $\color{RedOrange}\mathcal{L}$ indicates the leakage event and the qubit wire is colored red while it is leaked. The Pauli envelope is the composition of the indicated depolarizing error channels, {$\color{blue}\mathcal{D}$}.}
    \label{fig:depol-pauli-envelope-ex}
\end{figure}

For the depolarizing and tailored models of leakage effects, the Pauli envelope is simply the composition of the error channels applied to the qubits that the leaked qubit interacts with until it is reset, plus a depolarization on the leaked qubit itself when it is erasure-checked and reset to the logical subspace (for EC schedules 1, 2, and 4) or before it is measured (for EC schedule 8); see Fig.~\ref{fig:depol-pauli-envelope-ex} for an example. 
These final depolarizations account for the fact that the qubit is reinitialized to a random state when an erasure check flags a leakage, and that there is no logical information in a three-state measurement when the qubit is leaked. 

\begin{figure*}[htb]
    \centering
    \begin{quantikz}[row sep={6pt}, column sep={8pt}, align equals at=3]
& [-4pt] \rz& \targ{}\cancelgate[3]{} &                         &                        &                        &    & [-4pt] \\
& \rx       &                         & \ctrl{1}\cancelgate[2]{}&                        &                        &    & \\
& \leak{5}  & \ctrl{-2}\setwiretype{n}& \targ{}                 & \targ{}\cancelgate[2]{}& \targ{}\cancelgate[3]{}& \mzEC& \setwiretype{q} \\
& \rx       &                         &                         & \ctrl{-1}              &                        &    & \\
& \rz       &                         &                         &                        & \ctrl{-2}              &    & 
    \end{quantikz}
    $=$
    \begin{quantikz}[row sep={6pt}, column sep={8pt}, align equals at=3]
& [-4pt] \rz          & \targ{}  &                      &         &          &          &       &    & [-4pt]\\
& \rx                 &          &                      & \ctrl{1}&          &          &       &    & \\
& \errgate{\mathsf{r}}& \ctrl{-2}& \errgate{\mathsf{rx}}& \targ{} & \targ{}  & \targ{}  & \depol& \mzEC& \\
& \rx                 &          &                      &         & \ctrl{-1}&          &       &    & \\
& \rz                 &          &                      &         &          & \ctrl{-2}&       &    & 
    \end{quantikz}
    $\subseteq$
    \begin{quantikz}[row sep={6pt}, column sep={8pt}]
& [-4pt] \rz& \targ{}  &       &         &          &          &       &    & [-4pt] \\
& \rx       &          &       & \ctrl{1}&          &          &       &    & \\
& \depol    & \ctrl{-2}& \depol& \targ{} & \targ{}  & \targ{}  & \depol& \mzEC& \\
& \rx       &          &       &         & \ctrl{-1}&          &       &    & \\
& \rz       &          &       &         &          & \ctrl{-2}&       &    & 
    \end{quantikz}
    \caption{Example of a leakage event's Pauli envelope for the skip-gate leaked-qubit effect and EC schedule 8 on the moonwalking surface code. 
    See Fig.~\ref{fig:depol-pauli-envelope-ex} for description of notation. 
    The first two diagrams show the correspondence between skipped gates and erroneously inserted resets (these resets are given dashed outlines and red filling to indicated that they are errors). 
    The latter two diagrams represent that the Kraus operators describing the inserted resets are superpositions of a non-strict subset of the Kraus operators describing completely depolarizing channels at the same locations. 
    In practical terms, the possible impacts on detector and logical outcomes of resets are a subset of the possible impacts of completely depolarizing noise at the same locations. 
    Therefore, the Pauli envelope for the leakage event illustrated in the leftmost diagram is the composition of the depolarizing error channels in the rightmost diagram.
    }
    \label{fig:skipgate-pauli-envelope-ex}
\end{figure*}
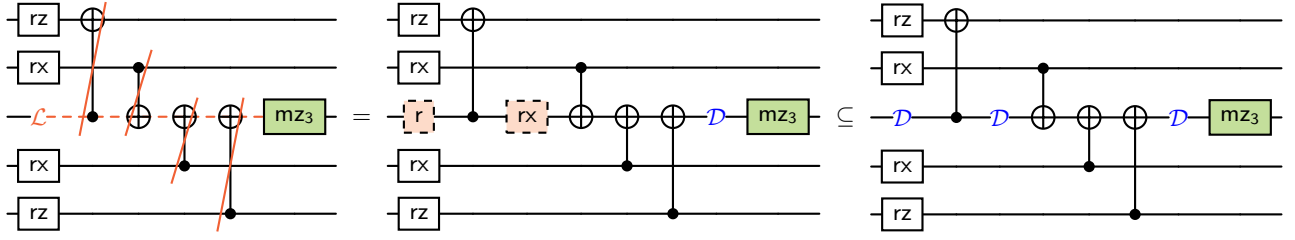

We now consider the skip-gate model of the leaked-qubit effect (see Sec.~\ref{sec:leaked-qubit-effects}). This effect is Clifford, but finding a leakage event's Pauli envelope aids in analysis and decoding. 
Notice that when the control qubit of a CX gate is in the $\ket{0}$ state, the target qubit is unaffected, and similarly, when the target qubit of a CX gate is in the $\ket{+}$ state, the control qubit is unaffected. Therefore, we can model the effect of skipped two-qubit gate by inserting resets before a leaked qubit would interact with another qubit.
As with the depolarizing and tailored leakage effect model, there is also a depolarization on the leaked qubit itself when it is erasure-checked and reset to the logical subspace (for EC schedules 1, 2, and 4) or before it is measured (for EC schedule 8). 
See the first two diagrams in Fig.~\ref{fig:skipgate-pauli-envelope-ex} for an example of the correspondence between skipped gates and qubit resets. 

The Kraus operators which implement $Z$-basis reset are ${K}_1=\ketbra{0}{0}=\frac{1}{2}(I+Z)$ and ${K}_2=\ketbra{0}{1}=\frac{1}{2}(X+iY)$, which includes all four Pauli operators with equal strength. 
The $X$-basis reset has similar Kraus operators.  
Consider replacing a reset with a full depolarization, $\mathcal{D}=\mathcal{E}\{I,X,Y,Z\}$. Clearly, the reset channel can be formed by superpositions of elements of the depolarizing channel. Therefore, if a decoder can correct a depolarization, it can correct a reset at the same location. 
In general, the error channel for leakages with the skip-gate effect consists of resets and depolarizations at particular circuit locations which depend on the leakage location. Call this channel $\mathcal{E}_\mathrm{exact}$; its Kraus operators are superpositions of Pauli products. 
For the purpose of decoding, we replace resets with full depolarizations to get an error channel $\mathcal{E}_\mathrm{decode}$, whose Kraus operators are Pauli products. The Kraus operators of $\mathcal{E}_\mathrm{exact}$ can be expressed as superpositions of the Kraus operators of $\mathcal{E}_\mathrm{decode}$. Therefore, $\mathcal{E}_\mathrm{decode}$ is a Pauli envelope for $\mathcal{E}_\mathrm{exact}$, and if our decoder can correct every error in $\mathcal{E}_\mathrm{decode}$, it can correct every error in $\mathcal{E}_\mathrm{exact}$. 
Usually, every error in $\mathcal{E}_\mathrm{decode}$ will be included in some error in $\mathcal{E}_\mathrm{exact}$, but coherence between resets will occasionally destructively interfere and remove some Pauli product entirely from $\mathcal{E}_\mathrm{exact}$. This is why $\mathcal{E}_\mathrm{decode}$ is an \emph{envelope} for $\mathcal{E}_\mathrm{exact}$---it has the building blocks for $\mathcal{E}_\mathrm{exact}$ but may contain extra, unused terms. It is a good choice of envelope because it usually does not contain more Kraus operators than necessary. 
See the latter two diagrams in Fig.~\ref{fig:skipgate-pauli-envelope-ex} for an example of the correspondence between reset and depolarization. 

In summary, when the leaked-qubit effect is skip-gates, a leakage event's Pauli envelope is the composition of fully depolarizing channels inserted at each of these circuit locations:
\begin{itemize}
    \item the circuit location of the leakage event
    \item wherever the leaked qubit changes between being a control or target of a CX
    \item where the leakage terminates.
\end{itemize} 
If the leakage is removed by a mid-circuit erasure check, then the Pauli envelope has a depolarization \emph{after} the erasure check. 
If the leakage is detected and removed by a three-state measurement, then the Pauli envelope has a depolarization \emph{before} the measurement. 
When a circuit uses CZ gates instead of CX gates, then a leakage event's Pauli envelope has a full depolarization wherever there is a Hadamard between two CZs. This substitutes for the depolarizations when a leaked qubit changes between being control or target of a CX. 

Notice that the number of depolarizations in the Pauli envelope for a set of skip-gate leakages will generally be a few times larger than the number of leakage locations. Therefore, we cannot say \textit{a priori} that a distance-$d$ surface code can correct $d-1$ erasures with infrequent erasure checks. 
Instead, we determined the number of correctable erasures by exhaustive search on distance-3 circuits; see Sec.~\ref{sec:min-weights}. 

Notice also that, for all leaked-qubit effects considered, the Pauli channels in the Pauli envelopes are either fully depolarizing ($\mathcal{D}$), fully dephasing ($\mathcal{E}\{I,Z\}$), or fully dephasing in the $X$-basis ($\mathcal{E}\{I,X\}$). 
Therefore, the DEM graph for a Pauli envelope contains only weight-zero edges. This will be relevant for quasi-MLE decoding (Appendix~\ref{app:mle-decoding}). 

\subsection{Marginal decoding}\label{app:marginal-decoding}

In most cases, we use only the marginal decoding strategy because it is fast.
It is identical to the effective circuit construction of \cite{guFaulttolerantQuantumArchitectures2025} for depolarizing leakage, but we formulate it in terms of DEM composition for computational efficiency and for easier extension to our quasi-MLE strategy described in the next subsection. 
In some circumstances---under the depolarizing and tailored leakage effect models, and under the skip-gate leaked-qubit effect with EC schedule 1---the marginal decoding strategy is equivalent to the quasi-MLE strategy.
This equivalence holds because in these cases the Pauli envelope for a later leakage event is a subset of the Pauli envelope for an earlier leakage event which could trigger a given erasure check. 
Furthermore, while the marginal decoding strategy deviates more significantly from the quasi-MLE solution under the skip-gate leaked-qubit effect with EC schedule 2, it preserves minimum fault weight relative to the quasi-MLE strategy.

We now define some terms that will be used for both decoding strategies. 
Given a triggered erasure check $\mathrm{EC}_i$, the set of leakage events/circuit locations that could have triggered it is $\{\mathrm{L}_{i,j}\}$. 
Note that these leakage events trigger $\mathrm{EC}_i$ and no other erasure check.  
Given our error model (Fig.~\ref{fig:error-model}), these circuit locations are immediately before each CX between the triggered erasure check and the prior erasure check or reset on the same qubit. The only exception is when using erasure check schedule 8 on the static surface code when we must account for the leakage-SWAP gate. 
The number of leakage events that could have triggered erasure check $\mathrm{EC}_i$ is $N_i= \left|\{\mathrm{L}_{i,j}\}\right|$. When using erasure check schedule $n$, we have $N_i=n$ for qubits in the bulk and $N_i<n$ for qubits around the edges. 
Let $G_{\mathrm{L}(i,j)}$ be the DEM graph for leakage event $\mathrm{L}_{i,j}$'s Pauli envelope (see Appendix~\ref{app:pauli-envelope}). 

Given a triggered erasure check $\mathrm{EC}_i$, the probability that leakage event $\mathrm{L}_{i,j}$ occurred is approximately $P_{i,j}= 1/N_i$ (see Footnote~\cite{Note4}).
We construct an average DEM graph for this erasure check using scalar multiplication, Eq.~\eqref{eq:dem-scalar-mult}, and disjoint addition, Eq.~\eqref{eq:dem-comb-disj}: 
\begin{equation}\label{eq:avg-ec-dem}
    {G}_{\mathrm{EC}(i)}=\bigoplus_{j=1}^{N_i} P_{i,j}G_{\mathrm{L}(i,j)}.
\end{equation}
When used for MWPM, this graph will \emph{approximately} enforce the disjointness of leakage events, i.e., the fact that the triggered erasure check is associated with exactly one leakage event. 

To see how the graph Eq.~\ref{eq:avg-ec-dem} approximately enforces disjointness,
consider an edge, $e_\mathrm{unique}$, which is present in only one of the graphs $G_{\mathrm{L}(i,j)}$. For example, there is a depolarization event at the earliest circuit location out of all possibilities $\{\mathrm{L}_{i,j}\}$ if and only if the qubit leaked at that earliest circuit location. 
Then, this edge's probability in the average DEM graph, $G_{\mathrm{EC}(i)}$, is $p_{\mathrm{EC}(i)}[e_\mathrm{unique}]=1/(2N_i)$, and its weight is correspondingly high. Recall that all edges in the Pauli envelope graphs $G_{\mathrm{L}(i,j)}$ have probability $1/2$ and weight $0$. 
At the other extreme, consider an edge, $e_\mathrm{all}$, which is present in every graph $G_{\mathrm{L}(i,j)}$ for the erasure check $\mathrm{EC}_i$. For example, no matter when the qubit leaked, there is a depolarization event before it is measured (for erasure check schedule 8) or after it is erasure-checked (all other schedules). 
Then, this edge's probability in the average DEM graph, $G_{\mathrm{EC}(i)}$, is $p_{\mathrm{EC}(i)}[e_\mathrm{all}]=N_i\times 1/(2N_i)=1/2$, and its weight is 0. 
This edge is ``free'' to include in an MWPM solution. 
In general, an MWPM solution on graph $G_{\mathrm{EC}(i)}$ is more likely to include edges which are compatible with multiple of the leakage events which could have triggered the erasure check $\mathrm{EC}_i$, and less likely to include edges which are compatible with fewer of the leakage events. 
However, this is only an issue of likelihood---the solution may still contain edges which are mutually incompatible, i.e., which represent errors that cannot have come from the same leakage event and therefore imply the qubit ``leaked'' twice. 

We then add the average erasure check DEMs, Eq.~\eqref{eq:avg-ec-dem}, independently to get the full erasure check DEM graph: 
\begin{equation}\label{eq:full-EC-DEM}
    \widetilde{G}_\mathrm{EC} = \sum_{i=1}^{M} {G}_{\mathrm{EC}(i)}. 
\end{equation}
Independent addition is appropriate in this case because the effect of one leaked qubit on detector and logical outcomes is statistically independent of another's.

To perform marginal decoding, we require the baseline DEM graph for the circuit with no leakage, $G_\mathrm{base}$, and the list of triggered erasure checks, $\{\mathrm{EC}_i\}$.
The marginal decoding graph is 
\begin{equation}\label{eq:marginal-DEM}
    \widetilde{G} = G_\mathrm{base}+\widetilde{G}_\mathrm{EC},
\end{equation}
where $\widetilde{G}_\mathrm{EC}$ is given in Eq.~\eqref{eq:full-EC-DEM}. 
Performing minimum-weight perfect matching on $\widetilde{G}$ completes the marginal decoding strategy.

As seen in Sec.~\ref{sec:min-weights}, the marginal decoding strategy may reduce the minimum number of leakage events which which can cause an uncorrectable error even with erasure information. This is because the marginal decoding strategy does not strictly enforce the disjointness of leakage events, i.e., the fact that the triggered erasure check is associated with exactly one leakage event. 
See Fig.~\ref{fig:moon-uncorr} in Appendix~\ref{app:t_E} for an example of this failure of the  marginal decoding strategy. 

\subsection{Quasi-MLE decoding}\label{app:mle-decoding}

We are motivated by the failure of the marginal decoding strategy to maintain the expected minimum fault weight for all EC schedules to develop a new strategy which strictly enforces the disjointness of leakage events. This new strategy is described in this section. 

This decoding strategy is \emph{quasi}-MLE because (1) DEM hypergraphs are approximately decomposed into independent $X$ and $Z$ graphs and (2) we approximate each leakage location which could trigger a given erasure check as equally likely. Neither approximation is distance-reducing in our setting---that is, they do not affect the minimum number of physical errors required to cause a logical error. 
We first describe a brute-force approach to solving the quasi-MLE problem and then our actual branch-and-bound algorithm. 

For the decoding problem, we are given a syndrome, $\bar{S}$, and a set of triggered erasure checks, $\left\{ \mathrm{EC}_i \right\}$. Let $M=\left|\left\{ \mathrm{EC}_i \right\}\right|$.
Each erasure check has a set of leakage locations which could have triggered it: $\left\{ \mathrm{L}_{i,j} \right\}$. Let $N_i= \left|\{\mathrm{L}_{i,j}\}\right|$.
Which leakage location triggered each erasure check is independent. 

The challenge of decoding with infrequent erasure checks is that each erasure check is associated with exactly one leakage event. We call this the \emph{disjointness} constraint because which specific leakage event caused a given erasure check to be triggered is described by a disjoint probability distribution. We also refer to this as the ``disjointness of leakage events.''

In Appendix~\ref{app:brute-force-mle}, we describe the quasi-MLE problem in terms suggesting a brute-force approach. With this understanding of the problem, we move on to our actual approach based on a branch-and-bound algorithm in Appendix~\ref{app:branch-bound-mle}.

\subsubsection{Brute force approach}\label{app:brute-force-mle}

In the brute-force approach to the quasi-MLE problem, we try every combination of leakage locations. There are $\prod_{i=1}^M N_i$ such combinations. 
For each combination, $\mathbf{j}=(j_1, j_2, \dots, j_M)$, we construct the brute-force candidate graph $G^\mathrm{(bf)}_{\mathbf{j}}$ by independently adding (see Eq.~\eqref{eq:dem-comb-indep}) the DEM graph for each leakage location's Pauli envelope, $G_{\mathrm{L}(i,{j_i})}$ to the baseline DEM graph, $G_\mathrm{base}$:
\begin{equation}\label{eq:bf-graph}
    G^\mathrm{(bf)}_{\mathbf{j}} = G_\mathrm{base} + \sum_{i=1}^{M} G_{\mathrm{L}(i,{j_i})}.
\end{equation}
We attempt to find a minimum-weight perfect matching for $\bar{S}$ on each candidate graph $G^\mathrm{(bf)}_\mathbf{j}$. Let the weight of the solution be $w_\mathbf{j}$. In cases where no solution exists, we say the solution has infinite weight. 
To account for the probability that $\mathbf{j}$ is the correct combination of leakage locations, we adjust the weights by $\Delta w_\mathbf{j}=\sum_{i=1}^M w(P_{i,j_i})$, where $P_{i,j}$ is the probability that erasure check $i$ was triggered by its $j$th possible leakage location. 
The solution to the quasi-MLE problem is the solution with the lowest final weight $w'_\mathbf{j}=w_\mathbf{j}+\Delta w_\mathbf{j}$.

In Appendix~\ref{app:pauli-envelope}, we noted that all edges in a Pauli envelope DEM $G_{\mathrm{L}(i,j)}$ have zero weight. Therefore, only edges which are in $G_\mathrm{base}$ but none of $\left\{ G_{\mathrm{L}(i,{j_i})} \right\}$ contribute to the weight of the MWPM solution $w_\mathbf{j}$. 
Under the approximation that all leakage locations for a given erasure check are equally probable, $P_{i,j}=1/N_i$~\cite{Note4}, all $\Delta w_\mathbf{j}$ are equal. 
When the erasure bias is infinite, i.e., there are no Pauli errors, the baseline DEM $G_\mathrm{base}$ has no edges, so for all candidate graphs $G_\mathbf{j}^\mathrm{(bf)}$ which have a matching for $\bar{S}$, that solution's weight is 0. 
Therefore, when the erasure bias is infinite, the quasi-MLE problem is solved by finding \emph{any} combination of leakage locations whose candidate graph $G_\mathbf{j}^\mathrm{(bf)}$ has a matching for $\bar{S}$. 

\subsubsection{Branch-and-bound decoder}\label{app:branch-bound-mle}

Using this observation that, under infinite erasure bias, the quasi-MLE problem is solved by finding any combination of leakage locations which admits a matching for $\bar{S}$, we move on to our actual algorithm for quasi-MLE decoding. 
This algorithm solves the quasi-MLE decoding problem exactly in the infinite bias case, which we will justify throughout this section. 
For finite erasure bias, we provide heuristic arguments for why we expect our algorithm to solve the quasi-MLE problem, though it may fail in particularly pathological cases. 

\begin{figure}
    \centering
    \resizebox{0.85\linewidth}{!}{
    \tikzset{
  rect/.style={
    rectangle, draw, thick,
    text width=5cm, align=center,
    inner sep=4pt
  },
  donerect/.style={
    rectangle, draw, thick,
    align=center,
    inner sep=6pt,
    fill=green!10
  },
  innerrect/.style={
    rectangle, draw, thick,
    text width=4.8cm, align=center,
    fill=white,
    inner sep=4pt
  },
  diam/.style={
    diamond, draw, thick,
    aspect=2.2,
    text width=2.2cm, align=center,
    inner sep=-2pt
  },
  arrow/.style={-{Stealth[length=7pt]}, thick},
  loopbox/.style={
    rectangle, sharp corners, draw, thick,
    fill=gray!12,
    inner sep=0pt,
    outer sep=0pt
  }
}

\begin{tikzpicture}[node distance=0.5cm]

  \node[rect, rounded corners=8pt, inner sep=6pt] (build1)
    {\textbf{(1)} Construct marginal graph by averaging over each erasure check's possible leakage events};

  \node[rect, below=of build1] (mwpm1)
    {Run MWPM};

  \node[rect, below=of mwpm1] (init)
    {Initialize list of candidate solutions with marginal solution};

  \node[rect, below=of init] (select)
    {Select lowest-weight candidate solution from list};

  \node[diam, below=of select] (valid)
    {\textbf{(2)} Is solution valid?};

  \node[donerect, rounded corners=8pt, right=0.8cm of valid] (done)
    {Done};

  \node[rect, below=of valid] (remove)
    {Remove this solution from list of candidate solutions};

  \node[rect, below=of remove] (find)
    {Find ``problematic'' erasure check which the solution implies leaked twice};

  \node[below=0.75cm of find, text width=5.2cm, align=center, font=\small]
    (looplabel) {For each leakage event which could have triggered this erasure check};

  \node[innerrect, below=0.25cm of looplabel] (constrain)
    {\textbf{(3)} Constrain erasure check to correspond to this leakage event only};

  \node[innerrect, below=of constrain] (build2)
    {\textbf{(1)} Construct partially constrained graph, including only chosen leakage events for constrained erasure checks, and averaging over leakage events for unconstrained erasure checks};

  \node[innerrect, below=of build2] (mwpm2)
    {Run MWPM};

  \node[innerrect, below=of mwpm2] (addsol)
    {Add solution to list of candidate solutions};

  \begin{scope}[on background layer]
    \node[loopbox, fit=(looplabel)(constrain)(build2)(mwpm2)(addsol)] (loopfit) {};
      \path let \p1 = (loopfit.north west), \p2 = (loopfit.south east) in
        coordinate (loopboxnw) at (\x1-8pt,\y1+4pt)
        coordinate (loopboxse) at (\x2+8pt,\y2-8pt);
      \coordinate (loopboxbottom) at ($(loopboxnw |- loopboxse)!0.5!(loopboxse)$);
      \draw[fill=gray!12, draw, thick, sharp corners] (loopboxnw) rectangle (loopboxse);
  \end{scope}

  \draw[arrow] (build1)    -- (mwpm1);
  \draw[arrow] (mwpm1)    -- (init);
  \draw[arrow] (init)     -- (select);
  \draw[arrow] (select)   -- (valid);

  \draw[arrow] (valid) -- node[above, font=\small] {yes} (done);

  \draw[arrow] (valid) -- node[right, font=\small] {no} (remove);
  \draw[arrow] (remove)   -- (find);
  \draw[arrow] (find)     -- ($(loopboxnw)!0.5!(loopboxnw -| loopboxse)$);

  \draw[arrow] (constrain)  -- (build2);
  \draw[arrow] (build2) -- (mwpm2);
  \draw[arrow] (mwpm2)      -- (addsol);

  \draw[arrow, rounded corners=8pt]
    (loopboxbottom) -- ++(0,-0.5)
    -- ++(-3.5, 0)
    |- (select.west);

\end{tikzpicture}
    }
\caption{
    Overview of our branch-and-bound quasi-MLE decoder. The \emph{branching} is the construction of one more partially constrained graph for each leakage location which could have triggered the problematic erasure check. The \emph{bounding} is the fact that adding constraints can only increase MWPM solution weight, which guarantees that the currently known lowest-weight solution is the lowest weight solution possible if the candidate solution tree were extended.
    }
\label{fig:branch-bound-flowchart}
\end{figure}
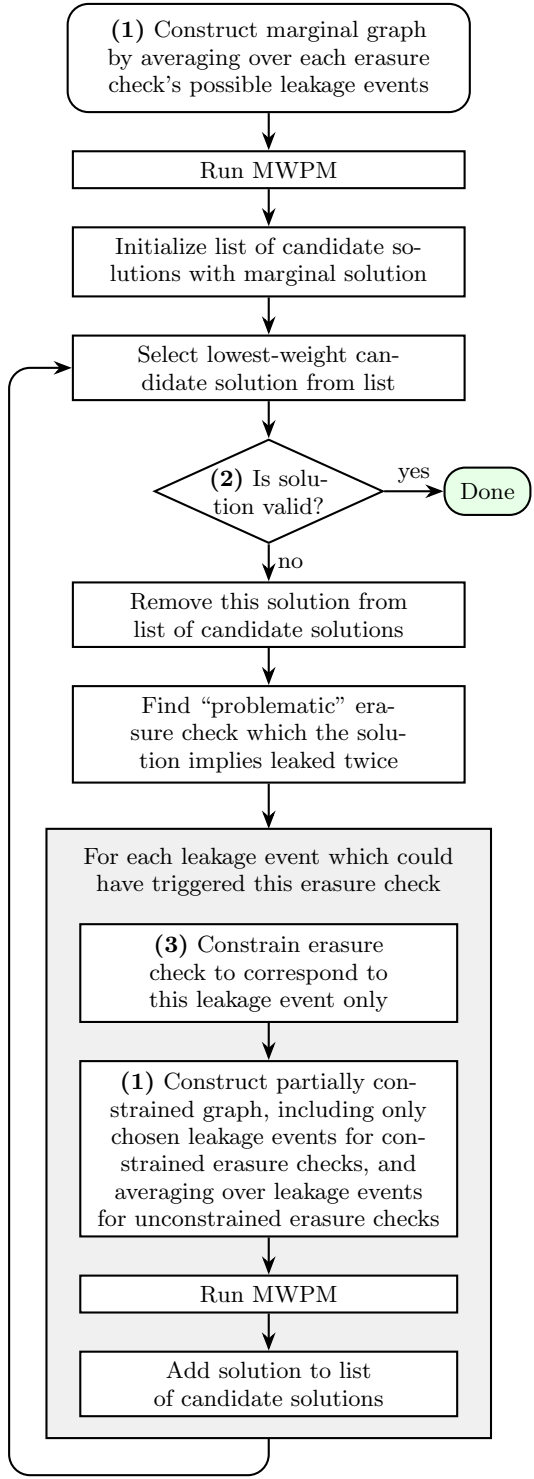

See Fig.~\ref{fig:branch-bound-flowchart} for an overview of our quasi-MLE decoding algorithm. 
There are three key operations which we describe in detail below: 
\begin{enumerate}
    \item Constructing a \emph{partially constrained} candidate graph and decoding on it
    \item Checking if a proposed solution is valid, i.e., if it is consistent with the disjointness constraint of leakage locations
    \item Further constraining the candidate graph based on inconsistencies with disjointness.
\end{enumerate}

\textbf{(1)}
Recall that there are $M=\left|\left\{ \mathrm{EC}_i \right\}\right|$ triggered erasure checks. 
A partially constrained candidate graph has $n\leq M$ erasure checks which are constrained. The constrained erasure checks have indices $\ibhat=(\ihat_1,\dots,\ihat_n)$, and for each of them we constrain ourselves to a particular leakage location which could have triggered that erasure check. The chosen leakage locations have indices $\jbhat=(\jhat_1,\dots,\jhat_n)$. 
The partially constrained candidate graph, $G^\mathrm{(pc)}_{\ibhat,\jbhat}$, is defined as
\begin{align}
    G^\mathrm{(pc)}_{\ibhat,\jbhat} &= 
        G_\mathrm{base} + 
        \widetilde{G}_{\mathrm{EC}\smallsetminus \ibhat} + 
        \widetilde{G}_{\mathrm{L}(\ibhat,\jbhat)}, \label{eq:pc-graph}\\
    \widetilde{G}_{\mathrm{EC}\smallsetminus \ibhat} 
        &= \sum_{i\notin\ibhat} {G}_{\mathrm{EC}(i)}
        = \sum_{i\notin\ibhat}\Big\{ \bigoplus_{j=1}^{N_i} P_{i,j}G_{\mathrm{L}(i,j)} \Big\}
        , \label{eq:partial-EC-graph} \\
    \widetilde{G}_{\mathrm{L}(\ibhat,\jbhat)} &=
    \sum_{k=1}^n P_{\ihat_k,\jhat_k} G_{\mathrm{L}(\ihat_k,\jhat_k)}. \label{eq:G-L-chosen}
\end{align}
See Eq.~\eqref{eq:dem-comb-indep}--\eqref{eq:dem-comb-disj} for definitions of DEM operations. 
The difference between $\widetilde{G}_{\mathrm{EC}\smallsetminus \ibhat}$ and the full erasure check marginal graph $\widetilde{G}_\mathrm{EC}$, Eq.~\eqref{eq:full-EC-DEM}, is its exclusion of the constrained erasure checks, 
$\big\{ \mathrm{EC}_{\ihat} : \ihat\in\ibhat \big\}$. 
We instead include the graphs for the chosen leakage locations, $\jbhat$, in $\widetilde{G}_{\mathrm{L}(\ibhat,\jbhat)}$. 
Adding a constraint $(\ihat', \jhat')$ to $G^\mathrm{(pc)}_{\ibhat,\jbhat}$ removes terms from the partially constrained graph, thereby decreasing edge probabilities and increasing the corresponding edge weights. 

When no erasure checks are constrained, the partially constrained graph is simply the marginal graph, Eq.~\eqref{eq:marginal-DEM}: ${n=|\ibhat|=0}\rightarrow G^\mathrm{(pc)}_{\ibhat,\jbhat}= G_\mathrm{base}+\widetilde{G}_{\mathrm{EC}\smallsetminus \ibhat}= G_\mathrm{base}+\widetilde{G}_{\mathrm{EC}} =\widetilde{G}$. 

When every erasure check is constrained, the partially constrained graph has the same edges as the brute-force candidate graph, $G^\mathrm{(bf)}_\mathbf{j}$, Eq.~\eqref{eq:bf-graph}, but the edges do not have the same weight because the partially constrained graph includes the probability of each leakage location $P_{i,j}$ in its definition. 
This difference is necessary because our branch-and-bound algorithm requires that adding constraints never decreases the MWPM solution weight. 
This difference cannot cause us to have the wrong solution in the case of infinite erasure bias because our task is satisfied by finding any solution consistent with the disjointness constraint. 
In the case of finite bias, it is unlikely for this difference to lead us to an incorrect solution because the weights of the edges contributed by leakage locations are much less than the weights of the edges contributed only by the baseline DEM, even though they are not identically zero as in the brute-force approach. Mistaken solutions become less and less likely as $p\rightarrow 0$. 

\begin{figure*}[tp]
    \centering
    \forestset{
    box/.style={draw, rounded corners=2pt, align=center, inner sep=2pt,},
    plain/.style={draw=none, align=center,},
    label tree/.style={
    for tree={tier/.option=level},
    level label/.style={
        before typesetting nodes={
        for nodewalk={current,tempcounta/.option=level,group={root,tree breadth-first},ancestors}{if={>OR={level}{tempcounta}}{before drawing tree={label me=##1}}{}},
        }
    },
    before drawing tree={
        tikz+={\coordinate (a) at ([xshift=-1cm]current bounding box.west);},
    },
    },
    label me/.style={tikz+={\node [anchor=mid east] at (a |- .center) {#1};}},
}

\begin{forest}
for tree={grow=south},
forked edges,
label tree,
for level=0{box},
for level=1{plain},
for level=3{box},
for level=4{plain},
for level=6{box},
[{$\ibhat=()$, $\jbhat=()$\\$G^\mathrm{(pc)}_{\ibhat,\jbhat}=\widetilde{G}$\\$w^\mathrm{(pc)}_{\ibhat,\jbhat}=1.0$}, fill=red!10, level label=no constraints,
    [{constrain $\mathrm{EC}_{5}$:\\let $\ihat_1=5$},
        [{let $\jhat_1=1$},
        [{$\ibhat=(5)$, $\jbhat=(1)$\\$w^\mathrm{(pc)}_{\ibhat,\jbhat}=1.0$}, fill=red!10, level label=single constraint,
            [{constrain $\mathrm{EC}_{8}$:\\let $\ihat_2=8$},
                [{let $\jhat_2=1$},
                [{$\ibhat=(5,8)$\\$\jbhat=(1,1)$\\$w^\mathrm{(pc)}_{\ibhat,\jbhat}=\infty$}, level label=two constraints]]
                [{let $\jhat_2=2$},
                [{$\ibhat=(5,8)$\\$\jbhat=(1,2)$\\$w^\mathrm{(pc)}_{\ibhat,\jbhat}=2.0$}]]
                [,,edge={draw=none}
                [{$\ldots$}, plain, edge={draw=none}]]
                [{let $\jhat_2=N_8$},
                [{$\ibhat=(5,8)$\\$\jbhat=(1,N_8)$\\$w^\mathrm{(pc)}_{\ibhat,\jbhat}=\infty$}]]
                [,phantom]
            ]
        ]]
        [{let $\jhat_1=2$},
        [{$\ibhat=(5)$, $\jbhat=(2)$\\$w^\mathrm{(pc)}_{\ibhat,\jbhat}=3.0$}]]
        [,,edge={draw=none} [{$\ldots$}, plain, edge={draw=none}]]
        [{let $\jhat_1=N_5$},
        [{$\ibhat=(5)$, $\jbhat=(N_5)$\\$w^\mathrm{(pc)}_{\ibhat,\jbhat}=1.5$}, fill=red!10,
            [{constrain $\mathrm{EC}_{3}$:\\let $\ihat_2=3$},
                [,phantom]
                [{let $\jhat_2=1$},
                [{$\ibhat=(5,3)$\\$\jbhat=(N_5,1)$\\$w^\mathrm{(pc)}_{\ibhat,\jbhat}=2.0$}]]
                [{let $\jhat_2=2$},
                [{$\ibhat=(5,3)$\\$\jbhat=(N_5,2)$\\$w^\mathrm{(pc)}_{\ibhat,\jbhat}=1.8$}, fill=green!10, double]]
                [,,edge={draw=none}
                [{$\ldots$}, plain, edge={draw=none}]]
                [{let $\jhat_2=N_3$},
                [{$\ibhat=(5,3)$\\$\jbhat=(N_5,N_3)$\\$w^\mathrm{(pc)}_{\ibhat,\jbhat}=5.0$}]]
            ]
        ]]
    ]
]
\end{forest}
\caption{
Example tree for branch-and-bound decoding algorithm.
Recall that the entries of $\ibhat$ index the constrained erasure checks and the entries of $\jbhat$ index the leakage locations we have chosen for those erasure checks. 
We start at the top with no constraints ($\ibhat=()$, $\jbhat=()$) and construct the partially constrained graph, $G^\mathrm{(pc)}_{\ibhat,\jbhat}$, which is equivalent to the marginal graph, $\widetilde{G}$.
After decoding on this graph, we check the solution's consistency with the disjointness constraint and see that it is not valid. This is indicated by the red fill on the box. 
Suppose that the ``problem'' edge is covered by the graph for erasure check 5. We therefore branch on $\mathrm{EC}_5$ to get singly constrained graphs with the constraints indicated. 
We decode on each graph $G^\mathrm{(pc)}_{\ibhat,\jbhat}$, and then check the validity of the lowest-weight solution, which has $\ibhat=(5)$, $\jbhat=(1)$ in this example. 
If the solution is still not valid (as indicated by the red fill), we branch again, getting the left cluster of doubly-constrained graphs. 
In this example, we suppose that none of the solutions with $\ibhat=(5,8)$, $\jbhat=(1,\jhat_2)$ have weight less than 1.5, so the lowest-weight candidate solution has the single constraint $\ibhat=(5)$, $\jbhat=(N_5)$. 
If that candidate solution is still not valid, we branch from it, getting the right cluster of doubly-constrained graphs. 
We check the candidate solution of unknown validity with the lowest weight, which in this example now corresponds to the constraints $\ibhat=(5,3)$, $\jbhat=(N_5,2)$. The algorithm finishes when it finds a valid solution, as indicated by the green fill and double outline. 
The validity check is the slowest part of the algorithm, so we only do it on the most promising candidate solution at any given point in time. Finding the MWPM solution for each partially constrained candidate graph is comparatively fast. }
\label{fig:branch-bound-tree}
\end{figure*}
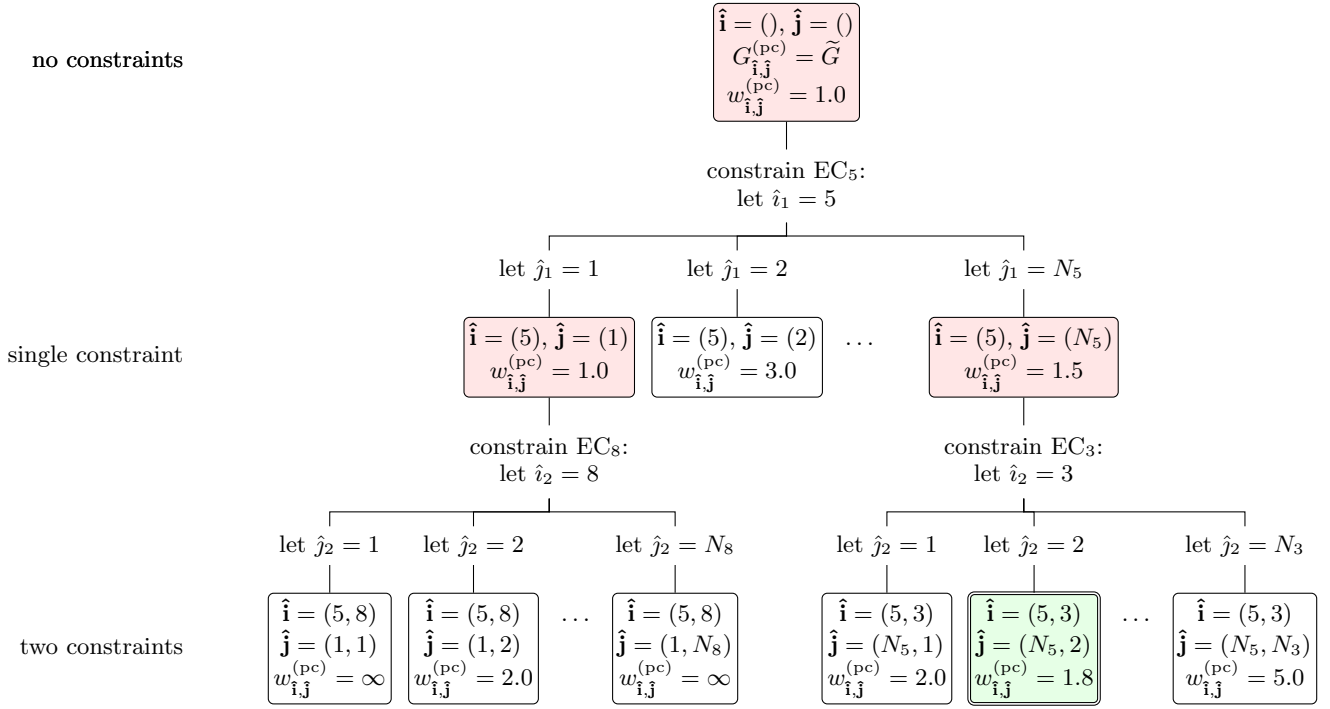

\textbf{(2)} If the partially constrained candidate graph $G^\mathrm{(pc)}_{\ibhat,\jbhat}$ has a matching for $\bar{S}$, then we can check if the MWPM solution is consistent with the disjointness constraint, i.e., if it is a valid solution. 
Let the matching have edges $\bar{E}$. 

Consider first the case where the erasure bias is infinite so the baseline graph, $G_\mathrm{base}$, has no edges. 
We wish to determine if there exists a combination of leakage locations, $\mathbf{j}$, one leakage location per triggered erasure check ($|\mathbf{j}|=M$), whose combined edges cover the edges in the matching, $\bar{E}$. 
That is, we want to know if $\exists \mathbf{j}\colon \bigcup_{i=1}^M E_{\mathrm{L}(i,j_i)}\supseteq \bar{E}$, where $E_x$ is the set of non-infinite-weight edges for the graph $G_x$. 
Because we have already constrained some erasure checks, we need only check those edges not covered by already-chosen leakage locations' DEMs, so the set of edges which actually need to be checked is $\bar{E}'=\bar{E} \setminus E_{\mathrm{L}(\ibhat,\jbhat)}$.
Furthermore, we need only consider the as-yet unconstrained erasure checks. 
This reduces the size of the problem depending on how many erasure checks have been constrained so far. 
In the case of infinite erasure bias, a matching on a partially constrained graph is valid if $\exists \mathbf{j}'\colon \bigcup_{i\notin \ibhat} E_{\mathrm{L}(i,j'_i)}\supseteq \bar{E}'$.

When the erasure bias is finite, the baseline DEM, $G_\mathrm{base}$, has finite-weight edges and its edges will generally be a superset of all leakage locations' DEMs' edges, so it may seem at first glance that any matching is valid. 
However, the solution's weight will be lower than it should be if it includes edges from disjoint leakage locations. One of these edges may have ``really'' come from a leakage event, and hence legitimately have low weight, but the other must have come from the baseline Pauli DEM, and should therefore have a higher weight. 
Therefore, checking the solution's consistency with the disjointness constraint is necessary to approach the true minimum-weight solution. 
As in the infinite-bias case, we need not check edges covered by the already-chosen leakage locations. 
We also do not need to check edges which are covered \emph{only} by the baseline DEM, because those have the full weight that they should. 
Therefore, the set of edges we need to check is $\bar{E}''=\bar{E}'\cap E_\mathrm{EC\smallsetminus\ibhat}$. 
For finite erasure bias, a matching on a partially constrained graph is valid if 
\begin{equation}\label{eq:validity-check}
    \exists \mathbf{j}''\colon \bigcup_{i\notin \ibhat} E_{\mathrm{L}(i,j''_i)}\supseteq \bar{E}''.
\end{equation}
When the erasure bias is infinite, $\bar{E}''=\bar{E}'$, so this formulation of the problem is general. 
This is a classic set-covering problem which can be solved by greedy algorithm with recursive backtracking. The algorithm is fast when a solution exists, which is true in the vast majority ($>99\%$ in the regime studied) of cases. 
Therefore, we can check if the proposed solution for the partially constrained graph is consistent with the disjointness of leakage locations. 
If the solution passes the validity check, the decoding is done due to bounding (see below). Otherwise, we move on to branching. 

\textbf{(3)} If a solution does not pass the validity check, the validity-checking algorithm returns the set of edges it was unable to cover. This set typically contains only one edge. We then identify an erasure check, $\mathrm{EC}_{i'}$, which has not yet been constrained, whose graph, $G_{\mathrm{EC}(i')}$, contains the problematic edge. 
The partially constrained graphs may be arranged in a tree, where at each level of the tree another constraint is added (see Fig.~\ref{fig:branch-bound-tree}). 
The \emph{branching} step is accomplished by adding child nodes to the node representing the partially constrained graph whose validity check just failed, $G^\mathrm{(pc)}_{\ibhat,\jbhat}$. 
There is one child node for each leakage location which could have triggered the erasure check $\mathrm{EC}_{i'}$. 
The child nodes have constraints $\ibhat'= (\ihat_1,\dots,\ihat_n,i')$ and $\jbhat'_k=(\jhat_1, \dots,\jhat_n,k)$. 
We construct the new partially constrained graphs, $G^\mathrm{(pc)}_{\ibhat',\jbhat'_k}$, and decode the syndrome on each of them. 
Let the weight of the MWPM solution on graph $G^\mathrm{(pc)}_{\ibhat,\jbhat}$ be $w^\mathrm{(pc)}_{\ibhat,\jbhat}$. When no solution exists, the weight is infinite. 
This concludes this branching step of the branch-and-bound algorithm. 

As stated when defining the partially constrained candidate graph, adding constraints can never reduce the weight of the solution. This is the \emph{``bound''} part of the branch-and-bound algorithm. 
It guarantees that all candidate solutions on the tree of candidate solutions have their weight lower-bounded by their parent node (see Fig.~\ref{fig:branch-bound-tree}). 
While our branch-and-bound decoding algorithm is running, there is a set of candidate solutions which we have determined are invalid and have branched from, and a set of candidate solutions which we have not yet checked the validity of. 
After the branching step is complete, we find the candidate solution of unknown validity with the lowest weight and check its validity. If the solution is valid, we accept it as our final solution. 
In the case of infinite erasure bias, this solution is guaranteed to be a solution to the quasi-MLE problem as explained in step 1. 
In the case of finite erasure bias, this solution is very likely to be a solution to the quasi-MLE problem, but it is not guaranteed. 
If the solution is not valid, we branch again. 

\section{Minimum-weight leakage faults}\label{app:t_E}

See main text Sec.~\ref{sec:min-weights} and Table~\ref{tab:erasure-t} for background. 
In this appendix, we present examples of minimum-weight leakage faults that can cause an \emph{uncorrectable} logical error even with erasure information, for situations where this minimum weight, $t_\mathrm{E}$, is less than the minimum number of leakage events that can cause an \emph{undetectable} logical error, $d_\mathrm{L}$. 
Both examples in this section use EC schedule 4 and the skip-gate leaked-qubit effect. 

\begin{figure}[!htb]
    \centering
    \includegraphics{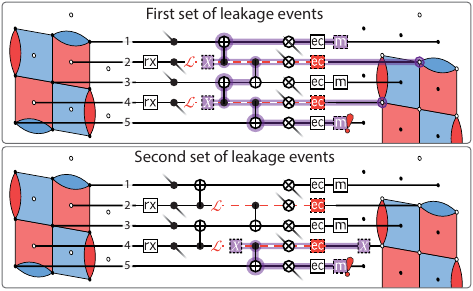}
    \caption{
    The first (upper) and second (lower) set of leakage events can cause the same syndrome and erasure checks but different logical outcomes in the walking surface code, showing that a distance-3 walking surface code circuit cannot always correct 2 leakage events even with erasure information. 
    Importantly, even though the upper and lower examples have different measurement outcomes, the detector outcomes, and therefore the syndromes, are the same. 
    This example uses EC schedule 4 and the skip-gate leaked-qubit effect. 
    The leakage event is indicated by $\color{RedOrange}\mathcal{L}$ and the qubit wire is dashed and colored red-orange while it is leaked. The erasure check ($\mathsf{ec}$) boxes have dashed outlines and are filled in red-orange when they are triggered. 
    The relevant effective Pauli errors (see Appendix~\ref{app:pauli-envelope}) are all $X$ errors and their propagation is shown in purple. Flipped measurement results have dashed outlines and are filled in purple. They are accompanied by red exclamation points when the flipped measurement also flips a detector. 
    The flipped measurement on qubit 1 in the upper panel does not trigger a detector because of the $X$ error which persists on qubit 2{---this is simply a result of the definition of detectors in the walking surface code and is not obvious by inspection}. 
    The second $X$ error on qubit 4 in the lower panel flips the code's logical $\bar{Z}$ operator.
    This $X$ error arises from the qubit being depolarized when it is reset to the logical subspace after the triggered erasure check.
    The initial and final states show detecting regions as described in Fig.~\ref{fig:moon-orig-reverse}.
    }
    \label{fig:orig-uncorr}
\end{figure}

Figure~\ref{fig:orig-uncorr} shows an instance where a distance-3 walking surface code circuit cannot correct 2 leakage events even with erasure information. Therefore, ${t_\mathrm{E}=\lceil d/2\rceil<d_\mathrm{L}=d}$ in this setting. 

\begin{figure}[!htb]
    \centering
    \includegraphics{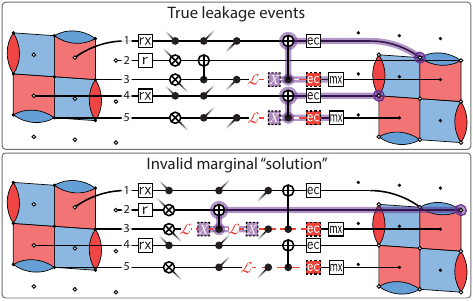}
    \caption{
    \label{fig:moon-uncorr}
    The marginal decoding strategy cannot always correct 2 leakage events in a distance-3 moonwalking surface code circuit with EC schedule 4 and the skip-gate leaked-qubit effect. 
    See Fig.~\ref{fig:orig-uncorr} for a description of notation. 
    In this example, the leakage events and effective Pauli errors illustrated in the upper panel are decoded as illustrated in the lower panel by the marginal decoder. Notice that the decoder assigns two $X$ errors to qubit 3 which would imply that it leaked twice. 
    The invalid marginal ``solution'' has lower weight than the true solution because the total marginal probability of those $X$ errors is higher than the total marginal probabilities of the true $X$ errors. 
    Specifically, the $X$ error before the second CNOT on qubit 3 can arise if qubit 3 leaks before either the first or second CNOT. 
    All other $X$ errors in these examples can only arise from a leakage event at the same time step as the effective Pauli error. 
    It is only by incorporating the leakage disjointness constraint that we know to discard the incorrect, lower-weight solution. 
    }
\end{figure}

Figure~\ref{fig:moon-uncorr} shows an instance where the marginal decoder cannot correct 2 leakage events, even with erasure information, on a distance-3 moonwalking surface code circuit. Therefore, maintaining ${t_\mathrm{E}=d_\mathrm{L}=d}$ in this setting requires quasi-MLE decoding. 

\clearpage

\end{document}